\documentclass[a4paper,11pt]{article}

\usepackage{jcappub} 
\usepackage{multirow,ulem}
\newcommand{\ish}[1]{{\color{blue}{#1}}}

\title{\boldmath Radial Oscillations of Hybrid Stars and Neutron Stars including Delta baryons: The Effect of a Slow Quark Phase Transition}

\author[a,b,1]{Ishfaq A. Rather \note{Corresponding author.},}
\affiliation[a]{Institut f\"{u}r Theoretische Physik, Goethe Universit\"{a}t, 60438 Frankfurt am Main, Germany}
\affiliation[b]{CENTRA, Instituto Superior T{\'e}cnico,
Universidade de Lisboa, 1049-001 Lisboa, Portugal}
\author[c]{Kauan D. Marquez,}
\affiliation[c]{CFisUC, Departmento de Física, Universidade de Coimbra, 3004-516 Coimbra, Portugal}
\author[d]{Betânia C. Backes,}
\affiliation[d]{School of Physics, Engineering and Technology, University of York, YO10-5DD York, United Kingdom}
\author[e]{Grigoris Panotopoulos,}
\affiliation[e]{Departamento de Ciencias F{\'i}sicas, Universidad de la Frontera, Casilla 54-D, 4811186 Temuco, Chile}
\author[b]{and Il{\'i}dio Lopes}
\emailAdd{rather@astro.uni-frankfurt.de, marquezkauan@gmail.com, betania.backes@york.ac.uk, grigorios.panotopoulos@ufrontera.cl, ilidio.lopes@tecnico.ulisboa.pt}

\abstract{
 We study radial oscillations of hybrid neutron stars composed of hadronic external layers followed by a quark matter core. We employ a density-dependent relativistic mean-field model including hyperons and ${\Delta}$ baryons to describe hadronic matter, and a density-dependent quark model for quark matter. We obtain the ten lowest eigenfrequencies and the corresponding oscillation functions of N, N+${\Delta}$, N+H, and N+H+${\Delta}$ equations-of-state with a phase transition to the quark matter at 1.4 and 1.8 ${M_{\odot}}$, focusing on the effects of a slow phase transition at the hadron-quark interface. We observe that the maximum mass is reached before the fundamental mode's frequency vanishes for slow phase transitions, suggesting that some stellar configurations with higher central densities than the maximum mass remain stable even when they undergo small radial perturbations. Future gravitational wave detectors and multi-messenger astronomy, complemented by robust microscopic models enabling exploration of various neutron star compositions, including hyperon content, are anticipated to impose precise limitations on the equation of state of baryonic matter under high-density conditions.
 }

\begin{document}
\maketitle
\flushbottom

\section{Introduction}
\label{sec:intro}

With radii of the order of 10 km and reaching up to two solar masses, neutron stars (NS) are made of the densest hadronic matter known to exist in the cosmos. As such, they are the ideal subject for studying cold, dense nuclear matter. From a theoretical perspective, the composition and macroscopic properties we can predict for these stars, such as their masses and radii, are determined by the nuclear equation of state (EoS). In addition to neutrons, NSs include a certain amount of protons to maintain the chemical equilibrium of the nuclear matter. Besides that, due to the nonperturbative nature of the strong interaction, we still know relatively little about the EoS of dense nuclear matter. Hence, the exact constitution of NSs is still unknown, particularly at densities significantly greater than nuclear saturation density (${n_0}$), where exotic degrees of freedom presumably exist. Consequently, confronting theoretical models with astrophysical observations is a powerful tool for extracting information about the true EoS of cold dense nuclear matter.

Theoretical models of neutron stars typically encompass the entire spin-1/2 baryon octet, which includes nucleons and hyperons \cite{Glendenning:1997wn}. The inclusion of hyperons in these descriptions arises from energy-related considerations. However, their incorporation also leads to a softening of the EoS, ultimately reducing the maximum mass achievable within a neutron star, leading to the so-called \textit{hyperon puzzle} \cite{2017hspp.confj1002B}.

In the relativistic mean-field framework, Glendenning \cite{1985ApJ...293..470G} initially explored various exotic degrees of freedom such as hyperons, kaons, and delta baryons (${\Delta}$) in the NS matter. Initially, the chosen coupling parameters suggested that $\Delta$ baryons might exist only at approximately 10 times the nuclear saturation density (${n_0}$) within these stars. However, recent research, guided by constraints from multiple experimental measurements on the coupling between $\Delta$ baryons and nucleons, indicates their potential presence within neutron stars \cite{PhysRevC.67.038801, LI2018234, RADUTA2021136070, PhysRevC.106.055801, PhysRevD.107.036011, PhysRevD.102.063008}. Furthermore, these studies suggest that ${\Delta}$ baryons could constitute a substantial portion of the baryonic content within NS matter, significantly influencing the stars' properties. Moreover, given that ${\Delta}$ baryons are only about 30\% heavier than nucleons (${m_{\Delta}}$ = 1232 MeV versus $m_{n}$ = 939.6 MeV) and lighter than the heaviest spin-1/2 baryons of the octet (the ${\Xi}$ hyperons, ${m_\Xi=1315}$ MeV), it is plausible to anticipate their existence within neutron stars at a density range similar to that of hyperons, approximately around 2-3$n_0$.

The recent achievement of detecting gravitational waves (GW) by the LIGO and Virgo Collaborations from the GW170817 event, a binary neutron star (BNS) merger \cite{PhysRevLett.119.161101, PhysRevLett.121.161101}, has significantly enhanced our ability to study dense matter properties under incredibly challenging conditions. These GW signals stemming from BNS mergers offer substantial insights, providing crucial constraints on both the EoS and the internal composition of neutron stars \cite{PhysRevLett.119.161101, PhysRevLett.121.161101, PhysRevLett.120.261103, PhysRevLett.120.172702, Capano2020}. Furthermore, the oscillations within neutron stars emit GWs across various frequency modes, enabling investigations into the star's internal constituents and a wide array of its properties \cite{PhysRevLett.77.4134, 10.1046/j.1365-8711.1998.01840.x}.

 Regarding excitation and detectability, numerous astrophysical mechanisms may excite oscillation modes of stars, such as tidal effects in binaries, starquakes caused by cracks during supernova explosions, magnetic reconfiguration, or any other form of instability \cite{Ex1,Ex2,Ex3,Ex4}. The Kepler and CoRoT missions have already measured the oscillation spectra of solar-like, white dwarf, and red giant stars \cite{corot1,corot2,corot3,Kepler}. Current wave detectors---such as LIGO---cannot detect the radial oscillations studied here due to their low sensitivity at the kHz frequency range. However, the third-generation ground-based gravitational wave detectors are expected to have a much higher sensitivity (by an order of magnitude), such as the Cosmic Explorer \cite{CE} and the Einstein Telescope \cite{ET}. These detections could give us information related to neutron star masses, frequencies, tidal Love numbers, amplitudes of the modes, damping times, and moments of inertia. Upon comparison between accumulated experimental and observational data (both from nuclear matter properties and multi-messenger astronomy) and well-accepted theoretical predictions, we should be able to probe and infer the equation of state of dense matter. For a recent work in which the authors studied the implications of the neutron star equation of state on stellar properties discussing composition and baryonic degrees of freedom as well, such as hyperon content, see e.g. \cite{Providencia:2023rxc}.

Following their formation in supernovae, oscillating neutron stars (NSs) emit a range of frequencies determined by the restoring force, and these oscillations could originate from various sources \cite{PhysRevLett.116.181101, PhysRevLett.108.011102, Chirenti_2017}. All in all, oscillations fall into two types: non-radial and radial.  Chandrasekhar \cite{PhysRevLett.12.114, 1964ApJ...140..417C} conducted groundbreaking research on the radial oscillations of stellar models, which provided important information about the stability and the EoS of compact stars. Thorne and Campolattaro \cite{1967ApJ...149..591T} investigated the non-radial oscillations of relativistic stars in the interim; these were subsequently named quasinormal modes (QNMs) because of the gradual reduction of the oscillations caused by gravitational wave (GW) emission. Typical nonrotating relativistic fluid stars are classified as QNMs according to their polar and axial properties. The polar modes include the fundamental ($f$), pressure ($p$), and gravity ($g$) modes. The axial modes contain just the spacetime ($w$) modes. Radial oscillations are hard to detect because they can't generate GWs by themselves. On the other hand, their correlation with non-radial oscillations increases GW signals and hence increases the likelihood of detection \cite{PhysRevD.73.084010,PhysRevD.75.084038}. Furthermore,  Chirenti et al. \cite{Chirenti_2019} noted that a hyper-massive NS and a brief gamma-ray burst are produced in the post-merger event of BNS and that these events may be impacted by radial oscillations. It could be possible to observe the hyper-massive NS's high-frequency oscillations in the range of 1-4 kHz. Since the radial $f$-mode oscillations of NSs depend on the EoS of compact stars, they are highly significant, as they could be detected in the best-case scenario by the current generation LIGO/Virgo/KAGRA detectors \cite{PhysRevLett.122.061104, 10.1093/ptep/ptac073}. Other possibilities include the third-generation detectors, like the Cosmic Explorer and the Einstein Telescope \cite{2019BAAS...51c.251S, Punturo_2010, 2021arXiv211106990K}. In many astrophysical circumstances, the radial $f$-mode frequency is also expected to be stimulated, leading to efficient GW emission. 

Similar to the various families of modes originating from distinct physical sources reported in \cite{kokkostas}, there are two families of radial oscillation modes that are essentially distinct from each other. The high-density core of the neutron star is home to one family, while the low-density envelope is home to the other \cite{1997A&A...325..217G}. A ``wall'' in the adiabatic index divides the two regions because of the drastic change in the matter's stiffness at the neutron drip point. This wall effect exists for any realistic EoS since it is associated with the neutron drip point---which is a feature of the low-pressure regime and is the same for all EoSs. 

In this work, we investigate radial oscillations of NSs with different hadronic matter compositions, such as hyperons and delta baryons, followed by a phase transition to quark matter. Numerous studies delved into the radial oscillations of neutron stars (NSs) featuring diverse exotic phases, such as dark matter and deconfined quark matter \cite{10.1093/mnras/stac2622, Routaray:2022acz, kokkostas, galaxies11020060, PhysRevD.101.063025, PhysRevD.98.083001, PhysRevD.107.103039, rather2023quark}. In our prior research \cite{PhysRevD.107.123022}, we focused on the radial oscillations of hadronic matter containing hyperons and ${\Delta}$ baryons. This present work extends that study by introducing a phase deconfinement transition to quark matter, exploring the interplay of hyperons, ${\Delta}$ baryons, and the transition process---a combined investigation yet to be addressed in the literature---with the additional feature of describing both hadronic and deconfined quark matter within the formalism of density-dependent relativistic mean-field models.

Our study focuses specifically on the fundamental ($f$) and pressure ($p$) modes for radial oscillations since a positive square of these kinds of eigenfrequency indicates that the star is stable against radial perturbations. Depending on the phase change in relation to the oscillation timescale, investigations of radial oscillations are typically divided into slow and rapid conversions. It has been shown that $\partial M/\partial n_c$ $>$ 0 is insufficient for determining radial stability for slow conversions in standard hybrid star settings. \cite{Pereira:2017rmp, Lugones:2021zsg}.
 Indeed, the assessment of static solutions with varying central densities ($n_c$) has traditionally been a reliable method to evaluate the stability of one-phase stars, typically characterized by unstable configurations displaying $\partial M/\partial n_c$ $<$ 0. However, this fundamental criterion does not always hold in certain scenarios, such as those involving charged strange quark stars, hybrid stars combining quark and hadronic phases, and color superconducting quark stars.
Research on multi-phase neutron stars holds considerable importance as it can unveil unique characteristics specific to these objects, offering potential observable indicators. Of particular interest is the eigenfrequency spectrum of these systems. This spectrum serves as a valuable tool for stability assessments of such multi-phase systems and has gained significance alongside the advancements in gravitational wave astronomy \cite{GW170817,LIGOGW170817,PhysRevLett.122.061104}. This significance stems from the potential for observational insights into the EoS of dense hadronic matter. Such observations could reveal the presence of exotic baryons, or indicate a transition from hadronic matter to a deconfined quark phase---known as a hadron-quark deconfinement phase transition. 
     
Our work is organized as follows: in Subsection (\ref{sec:eos1}), the EoS for the DD-RMF model along with the addition of $\Delta$ baryons and the couplings used is discussed. Subsections (\ref{sec:eos1}) and (\ref{sec:hyb}) discuss the deconfined quark matter model used for the study and the construction of phase transition/hybrid EoS, respectively. The TOV and the Sturm-Liouville
eigenvalue equations for the internal structure and radial
oscillations of NSs are introduced in Subsection (\ref{nsprop}) and (\ref{radial}), respectively. Subsection (\ref{junction}) discusses the necessary junction conditions. In Section (\ref{results}), the EoS and the Mass-Radius profile for different compositions of the matter are discussed in Subsection (\ref{mr}). Subsection (\ref{profile}) describes the numerical results obtained for NS EoS with hyperons and $\Delta$ baryons followed by a phase transition to the quark matter. The summary and concluding remarks are finally given in Section (\ref{summary}).

\section{Theoretical framework: Microphysics}

\subsection{Hadronic Matter}\label{sec:eos1}

From the plethora of RMF models available in the literature, less than 40 out of 263 were shown to satisfy nuclear bulk properties completely \cite{PhysRevC.90.055203}. Notably, some of the most successful models in doing so are density-dependent relativistic mean-field (DD-RMF) models. This study employs a modified version of the $\sigma-\omega$ hadrodynamics model to characterize hadronic matter. This model simulates the strong interaction among hadrons through virtual meson exchanges. In the density-dependent variant used here, the coupling between baryons and mesons adjusts with the density. The DD-RMF models provide a widely utilized and successful framework to NS matter description \cite{PhysRevLett.68.3408}, replacing the self- and cross-coupling of various mesons in the RMF model with density-dependent nucleon-meson coupling constants. This approach yields results comparable to other models and ensures uniform assessment of neutron star (NS) properties. Moreover, it incorporates characteristics from the Dirac-Brueckner model, utilizing microscopic interactions at varying densities as foundational input. Parameter sets such as such as DDME1 \cite{PhysRevC.66.024306}, DDME2 \cite{PhysRevC.71.024312}, and DDMEX  \cite{TANINAH2020135065}, exhibit consistency in reproducing nuclear matter's experimental properties even under astrophysical constraints \cite{dutra2014,lourencco2019,malik2022}, and yield a very stiff equation of state (EoS), predicting massive NSs with maximum masses typically falling within the range of 2.3-2.5$M_{\odot}$ \cite{Rather:2020lsg, PhysRevC.103.055814}.

The DD-RMF parameterization adopted here considers the scalar meson $\sigma$, the vector mesons $\omega$ and $\phi$ (that carries hidden strangeness), isoscalars,  and the isovector-vector meson $\vec\rho$, with the meson couplings depending on the baryonic density $n_b$ as
\begin{equation}
    g_{i b} (n_B) = g_{ib} (n_0) \frac{a_i +b_i (\eta + d_i)^2}{a_i +c_i (\eta + d_i)^2} 
\end{equation}
for $i=\sigma, \omega, \phi$ and 
\begin{equation}
    g_{\rho b} (n_B) = g_{ib} (n_0) \exp\left[ - a_\rho \big( \eta -1 \big) \right],
\end{equation}
for $i=\rho$, where $\eta =n_B/n_0$ and $n_0$ is the saturation density \cite{ddme2}. 

All thermodynamic quantities can be calculated in the standard way for RMF models. 
 The baryonic and scalar densities of a baryon of the species $b$ are given, respectively, by
\begin{equation}
n_b = \frac{\gamma_b}{2\pi ^{2}}\int_{0}^{{k_F}_b}dk\, k^{2}=\frac{\gamma_b}{6\pi ^{2}}{k_F}_b^{3}, \label{eq:rhobarion}
\end{equation} 
and
\begin{equation}
 n^s_b=\frac{\gamma_b}{2\pi ^{2}}\int_{0}^{{k_F}_b} dk \frac{k^{2}m_b^\ast}{\sqrt{k^{2}+{m_b^\ast}^{2}}}, \label{eq:rhoscalar}
\end{equation} 
with ${k_F}$ denoting the Fermi momentum,  since we assume the stellar  matter to be at zero temperature, and $\gamma_b$ is the spin degeneracy factor
The effective masses are 
\begin{equation}
    m_b^\ast =m_b- g_{\sigma b} \sigma_0 .
\end{equation}

The energy density is given by
\begin{align}\label{1a}
\mathcal{E}={}& \sum_b \frac{\gamma_b}{2\pi^2}\int_0^{{k_{F}}_b} dk k^2 \sqrt{k^2 + {m_b^\ast}^2}+ \frac{m_\sigma^2}{2} \sigma_0^2+\frac{m_\omega^2}{2} \omega_0^2 +\frac{m_\phi^2}{2} \phi_0^2  + \frac{m_\rho^2}{2} \rho_{03}^2 ,
\end{align}
where the sum runs over the baryons considered in a given matter composition (nucleons $N=\{n,p\}$, hyperons $H=\{\Lambda,\Sigma^-,\Sigma^0,\Sigma^+,\Xi^-,\Xi^0\}$), and/or particles of the spin-3/2 baryon decuplet such as the deltas $\Delta=\{\Delta^-,\Delta^0,\Delta^+,\Delta^{++}\}$. The pressure is given by the fundamental relation
\begin{equation}
    P =\sum_i \mu_i n_i - \mathcal{E} + n_B \Sigma^r,
\end{equation}
which receives a correction from the rearrangement term due to the density-dependent couplings, to guarantee thermodynamic consistency and energy-momentum conservation~\cite{Typel1999, PhysRevC.52.3043},
\begin{equation}
    \Sigma^r = \sum_b \Bigg[ \frac{\partial g_{\omega b}}{\partial n_b} \omega_0 n_b + \frac{\partial g_{\rho b}}{\partial n_b} \rho_{03} I_{3b}  n_b+ \frac{\partial g_{\phi b}}{\partial n_b} \phi_0 n_b  - \frac{\partial g_{\sigma b}}{\partial n_b} \sigma_0 n_b^s\Bigg]. \label{rearr1}
\end{equation}
The effective chemical potentials read
\begin{equation}
      \mu_b^\ast = \mu_b- g_{\omega b} \omega_0 - g_{\rho b} I_{3b} \rho_{03} - g_{\phi b} \phi_0 - \Sigma^r,
\end{equation}
where the $\mu_b$ are determined by the chemical equilibrium condition
\begin{equation}
    \mu_b=\mu_n-q_b\mu_e, \label{beta}
\end{equation}
in terms of the chemical potential of the neutron and the electron, with $\mu_\mu=\mu_e$. The particle populations of each individual species are determined by Eq.~\eqref{beta} together with the charge neutrality condition $\sum_i n_iq_i=0$, where $q_i$ is the charge of the baryon or lepton $i$. Leptons are admixed in the hadronic matter as a free non-interacting fermion gas ($\lambda=\{e,\mu\}$), as their inclusion is necessary in order to ensure the $\beta$-equilibrium and charge neutrality essential to stellar matter.

\begin{table}[!ht]
\centering
		\caption{DD-RMF parameters considered in this work (top) and its predictions for the nuclear matter properties at saturation density (bottom).\label{T1} }
\begin{tabular}{ c c c c c c c }
\hline
$i$ & $m_i(\text{MeV})$ & $a_i$ & $b_i$ & $c_i$ & $d_i$ & $g_{i N} (n_0)$\\
 \hline
 $\sigma$ & 550.1238 & 1.3881 & 1.0943 & 1.7057 & 0.4421 & 10.5396 \\  
 $\omega$ & 783 & 1.3892 & 0.9240 & 1.4620 & 0.4775 & 13.0189  \\
 $\rho$ & 763 & 0.5647 & --- & --- & --- & 7.3672 \\
 \hline
\end{tabular}

\vspace{10pt}

\begin{tabular}{c|cc}
\hline 
Quantity & Constraints \cite{dutra2014, Oertel:2016bki} & This model\\\hline
$n_0$ ($fm^{-3}$) & 0.148--0.170 & 0.152 \\
 $-B/A$ (MeV) & 15.8--16.5  & 16.4  \\ 
$K_0$ (MeV)& 220--260   &  252  \\
 $S_0$ (MeV) & 31.2--35.0 &  32.3  \\
$L_0$ (MeV) & 38--67 & 51\\
\hline
\end{tabular}
\end{table}

\begin{table}[!ht]
\centering
\caption {Baryon-meson coupling constants  $\chi_{ib}$ \cite{Lopes1,issifu}.
\label{T2}}
\begin{tabular}{ c c c c c } 
\hline
 b & $\chi_{\omega b}$ & $\chi_{\sigma b}$ & $I_{3b}\chi_{\rho b}$ & $\chi_{\phi b}$  \\
 \hline
 $\Lambda$ & 2/3 & 0.611 & 0 & 0.471  \\  
  $\Sigma^{-}$,$\Sigma^0$, $\Sigma^{+}$ & 2/3 & 0.467 & $-1$, 0, 1 & -0.471 \\
$\Xi^-$, $\Xi^0$  & 1/3 & 0.284 & $-1/2$, 1/2 & -0.314 \\
  $\Delta^-$, $\Delta^0$, $\Delta^+$, $\Delta^{++}$   & 1 & 1.053 & $-3/2$, $-1/2$, 1/2, 3/2 & 0  \\
  \hline
\end{tabular}
\end{table}

The model parameters are fitted from experimental constraints of nuclear matter at or around the saturation density $n_0$, namely the binding energy $B/A$, compressibility modulus $K_0$, symmetry energy $S_0$, and its slope $L_0$, shown in Table~\ref{T1} \cite{Typel1999, PhysRevC.52.3043}.
These parameters are fitted for nucleonic (protons and neutrons only) matter, and to determine the meson couplings to other hadronic species one must define the ratio of the baryon coupling to the nucleon one as $\chi_{ib}=g_{i b}/g_{i N}$, with $i = \{\sigma,\omega,\phi,\rho\}$. In this work, we consider hyperons and/or deltas admixed in the nucleonic matter and
 follow the proposal of ~\cite{Lopes1} to determine the $\chi_{ib}$ ratios through symmetry arguments, such as the fact that the Yukawa couplings terms present in the Lagrangian density of the DD-RMF models must be invariant under SU(3) and SU(6) group transformations. Hence, the couplings can be fixed to reproduce the potentials  $U_\Lambda =-28$~MeV, $U_\Sigma= 30$~MeV, $U_\Xi=-4$~MeV and $U_\Delta= -98$~MeV in terms of a single free parameter $\alpha_V$. Our choice of $\alpha_V=1.0$ for the baryon-meson coupling scheme corresponds to an unbroken SU(6) symmetry, and the values of $\chi_{ib}$ are shown in Table~\ref{T2}.

\subsection{Deconfined Quark Matter}\label{sec:eos2}
In this work, we chose the density-dependent quark mass (DDQM) model to describe quark matter. It is a simple and versatile model that has been utilized to study the deconfinement phase transition \cite{backes2021effects}, making it a good fit for our investigation of hybrid stars. Within this model, we emulate confinement by the density dependence introduced in the quark masses as
\begin{equation}
    m_i = m_{i0} + \frac{D}{n_B^{1/3}} + Cn_B^{1/3} = m_{i0} + m_I,
    \label{masses}
\end{equation}
where ${m_{i0}}$ (${i = u, d, s}$) is the current mass of the $i$th quark, ${n_B}$ is the baryon number density and ${m_I}$ is the density dependent term that encompasses the interaction between quarks. The model has two free parameters: ${D}$, that dictates linear confinement; and ${C}$, that is responsible for leading-order perturbative interactions \cite{Xia2014}.

Whenever we introduce a new dependence on the state variables---such as density, temperature, or magnetic field---, we need to carefully account for the issue of thermodynamic consistency, similarly to relation \eqref{rearr1} for the DD-RMF model. To overcome this problem, we follow the formalism of \cite{Xia2014}, that presents a thermodynamically consistent DDQM model. At zero temperature, the differential fundamental relation reads
\begin{equation}
    \text{d}\mathcal{E} = \sum_i \mu_i\text{d}n_i,
    \label{diff-fundamental-eq}
\end{equation}
where ${\mathcal{E}}$ is the matter contribution to the energy density of the system, ${\mu_i}$ are the particles' chemical potentials and ${n_i}$ are the particle densities. 

We then introduce effective chemical potentials to the model. Under this perspective, the energy density can be viewed as the one for a free system with particle masses ${m_i(n_B)}$  and effective chemical potentials ${\mu_i^*}$
\begin{equation}
    \mathcal{E} = \Omega_0 (\{\mu_i^*\},\{m_i\}) + \sum_i \mu_i^* n_i,
    \label{free-system-fundamental-eq}
\end{equation}
where ${\Omega_0}$ is the thermodynamic potential of a free system. The differential form of Eq. \eqref{free-system-fundamental-eq} is
\begin{equation}
    \text{d}\mathcal{E} = \text{d}\Omega_0 + \sum_i \mu_i^* \text{d} n_i + \sum_i n_i \text{d}\mu_i^*.
    \label{initial-diff-free-system}
\end{equation}
Explicitly, we can write d${\Omega_0}$ as
\begin{equation}
    \text{d} \Omega_0 = \sum_i \frac{\partial \Omega_0}{\partial \mu_i^*} \text{d}\mu_i^* + \sum_i \frac{\partial \Omega_0}{\partial m_i}\text{d}m_i
\end{equation}
with
\begin{equation}
    \text{d}m_i = \sum_j \frac{\partial m_i}{\partial n_j} \text{d}n_j,
\end{equation}
where the densities are connected to the effective chemical potentials by
\begin{equation}
    n_i = -\frac{\partial \Omega_0}{\partial \mu_i^*}
\end{equation}
to ensure thermodynamic consistency. Eq. \eqref{initial-diff-free-system} can then be rewritten as
\begin{equation}
    \text{d}\mathcal{E} = \sum_i \left(\mu_i^* + \sum_j \frac{\partial \Omega_0}{\partial m_j} \frac{\partial m_j}{\partial n_i}\right) \text{d}n_i,
    \label{diff-free-system}
\end{equation}
which should be consistent with the fundamental equation. Finally, by comparing Eqs. \eqref{diff-fundamental-eq} and \eqref{diff-free-system} one gets the relation between the real and the effective chemical potentials
\begin{equation}
    \mu_i = \mu_i^* + \sum_j \frac{\partial \Omega_0}{\partial m_j} \frac{\partial m_j}{\partial n_i}.
\end{equation}
It immediately follows that the pressure, ${P}$, is then given by

\begin{align}
    P ={}& -\mathcal{E} + \sum_i \mu_i n_i\\
    ={}& -\Omega_0 + \sum_i (\mu_i - \mu_i^*)n_i\nonumber\\
    ={}& -\Omega_0 + \sum_{i,j} \frac{\partial \Omega_0}{\partial m_j}n_i\frac{\partial m_j}{\partial n_i},
    \label{pressure-quarks}
\end{align}
which is the thermodynamically consistent EoS.

\subsection{Phase Coexistence and the Hybrid Equation of State}\label{sec:hyb}

The hadron-quark deconfinement transition is assumed to be a first-order phase transition at the high-density region of the QCD phase diagram, as foreseen by effective models. The thermodynamic description of this kind of process can be obtained by matching the equations of state of the two phases, determining the phase coexistence point. For the present study, we employ the Maxwell construction, which results in a set of hybrid EoS with a first-order phase transition that occurs at a critical value of the baryonic chemical potential and pressure.According to the Gibbs criteria, the point where
\begin{equation}
\begin{gathered}
  P^{(i)}=P^{(f)}=P_0,\\
  \mu^{(i)}(P_0)=\mu^{(f)}(P_0)=\mu_0,
\end{gathered} \label{eq:gibbscon}
\end{equation}
sets the transition between the initial (${i}$) and final (${f}$) homogeneous phases, both at ${T=0}$ MeV, with
\begin{equation}
 \mu^{ (i,f)}=\frac{\mathcal{E}^{(i,f)}+P^{(i,f)}}{n_B^{(i,f)}},
\end{equation}
where ${\mathcal{E}^{(i,f)}}$, ${P^{(i,f)}}$ and ${n_B^{(i,f)}}$ are the total energy density, pressure, and baryon number density, obtained from the effective models of each phase.
The conditions above leave the values of $P_0$ and $\mu_0$ to be determined from the equations of state of both hadronic and deconfined quark phases. 

For a given baryonic constitution of the hadron phase, the transition point is highly impacted by the choice of DDQM model free parameters\cite{backes2021effects}. Since the quark EoS is not well constrained by empirical measurements, this choice usually relies on observing the stability window according to the Bodmer-Witten hypothesis. The basis of this hypothesis is that when a system is composed of similar quantities of $u$, $d$, and $s$ quarks, a deconfined quark system becomes more bound than the confined one. Hence, hadronic matter would be metastable, and the ground state of nuclear matter would be deconfined quark matter---the so-called strange quark matter \cite{Bodmer1971,Witten1984}. The main implication of this conjecture to nuclear astrophysics is that if the core of a NS is composed of strange matter, the entire star converts into a strange star, composed solely of strange quark matter. In this work, our analysis focuses on hybrid stars---which are good candidates for massive NSs, as suggested by observational data \cite{Annala_2020}. Hence, the first constraint we impose on the DDQM free parameters is that they must not satisfy the Bodmer-Witten hypothesis.
Naturally, parameter sets that are well suited for describing strange stars are mutually exclusive with the ones for hybrid stars. However, studies of strange stars are still a powerful source of insight for good parameter choices. It is intuitive for the binding energy of quark matter to be comparable to the one of hadronic matter, thus it is sensible to restrict our study to parameter sets close to the stability window. We refer to Ref. \cite{Backes_2021} for a detailed investigation of the DDQM parameters effects on the stability window of the strange matter.

Lastly, we may look at the density profiles of strange stars described by the DDQM model. Previous studies found that, for large values of the ${C}$ parameter, the surface density of strange stars becomes comparable to or even below the one of nuclear saturation density \cite{Xia2014}. This effect points towards a phase transition, making such parameters a good fit for the description of hybrid stars. Moreover, an analysis of the hadron-quark phase transition from distinct effective models suggested that large values of the perturbative parameter lead to the phase transition occurring in densities higher than the one of nuclear saturation \cite{Backes_2021}. These factors, paired with such parameters producing strange stars with masses in the order of ${2 M_{\odot}}$, led to the choice of EoSs with high ${C}$ values for our analysis.

\begin{table}[!htb]
    \centering
\begin{tabular}{cc|cccc}
\hline
                   Composition  &      (${C, D^{1/2}}$)      & $\mu_0$ (MeV) & $P_0$  (MeV fm$^{-3}$)                & $n_B^{(H)}$ (fm$^{-3}$) & $n_B^{(Q)}$ (fm$^{-3}$) \\ \hline
\multirow{4}{*}{N}   & (0.65, 133) & \multicolumn{4}{c}{No Coexistence Point}                                                                      \\
                     & (0.65, 135) & 1205.91            & 68.01              & 0.387                  & 0.427                  \\
                     & (0.90, 125) & 1478.54 & 195.83 & 0.555    & 0.738    \\
                     & (0.90, 127) & 1500.39 & 208.01 & 0.567    & 0.773    \\ \hline
\multirow{4}{*}{N+$\Delta$}  & (0.65, 133) & 1392.95 & 181.38 & 0.597    & 0.759    \\
                     & (0.65, 135) & 1418.20 & 196.69 & 0.614    & 0.802    \\
                     & (0.90, 125) & 1669.20 & 373.42 & 0.793    & 1.141    \\
                     & (0.90, 127) & 1684.74 & 385.84 & 0.804    & 1.170     \\ \hline
\multirow{4}{*}{N+H}  & (0.65, 133) & 1297.08 & 117.40 & 0.555    & 0.580    \\
                     & (0.65, 135) & 1414.08 & 193.39 & 0.707    & 0.797    \\
                     & (0.90, 125) & \multicolumn{4}{c}{No Coexistence Point}                                                                      \\
                     & (0.90, 127) & \multicolumn{4}{c}{No Coexistence Point}                                                                      \\ \hline
\multirow{4}{*}{N+H+$\Delta$} & (0.65, 133) & 1524.62 & 299.79 & 0.862    & 1.046     \\
                     & (0.65, 135) & 1550.57 & 322.55 & 0.891    & 1.105    \\
                     & (0.90, 125) & \multicolumn{4}{c}{No Coexistence Point}                                                                      \\
                     & (0.90, 127) & \multicolumn{4}{c}{No Coexistence Point}                                                                      \\ \hline
\end{tabular}
    \caption{Phase coexistence points ($\mu_0,P_0$) according to the Gibbs criteria \eqref{eq:gibbscon}, for the hadron-quark deconfinement transition considering distinct baryonic constitutions of the hadron phase and DDQM parameter choices. The baryon number density at the transition for each phase is given by $n_B^{(H,Q)}$.}
    \label{tab:coex}
\end{table}

For this work, we obtained the EoS for the quark model \cite{betania-github} using quark current masses compatible with experimental data \cite{Workman:2022ynf}. We then selected some parameter sets which we deemed more suited for studying hybrid stars following the above-established criteria, and found their respective phase coexistence points with each baryonic constitution of the hadron phase, as illustrated in Table \ref{tab:coex}. We selected two sets of DDQM parameters to have a quark core with all the hadronic configurations, as with a single parameter set we observed either no crossing or a crossing at densities higher than the NS central density. It is clear that the parameter $C$ has a stronger effect in determining the existence and, if so, the placement of the coexistence point in the $\mu$-$P$ plane, with the parameter $D^{1/2}$ having a lower order fine-tuning impact. We favor the hybrid EoS that guarantees the dynamical stability of the branch to the left of maximum mass (as discussed in the following section), so our further analysis will be limited to the parameter sets (${C, D^{1/2}}$) = (0.90, 125 MeV) and  (${C, D^{1/2}}$) = (0.65, 133 MeV),  for the sake of concision.

 It is worth pointing out that the criteria  given by Eq. \eqref{eq:gibbscon} may be satisfied more than once for a given DDQM parameter choice. This point is disregarded as nonphysical, as it would indicate a transition from a stable deconfined quark phase at low densities into a hadronic phase at higher densities. As the quark matter EoS \textit{by construction} does not fulfill the Bodmer-Witten hypothesis, it is not possible consider the speculative ``inverted hybrid star'' configuration in this study (we refer the interested reader to \cite{Zhang:2023zth}). The second coexistence point corresponds to the hadron-quark deconfinement phase transition, which we then utilize for building the hybrid EoS.

\section{Theoretical framework: General Relativity}

\subsection{Structure equations}
\label{nsprop}

Because of the intense gravitational field, the structure and dynamics of NSs are governed by Einstein's equations of general relativity as 
\begin{equation}
    G_{\mu \nu} = R_{\mu \nu} - \frac{1}{2} R g_{\mu \nu} = 8\pi T_{\mu \nu}, 
\end{equation}
where $R_{\mu \nu}$ and $R$ represent the Ricci tensor and Ricci scalar, respectively. The gravitational constant $G$ is set equal to 1. The energy-momentum tensor expression, $T_{\mu \nu}$, is defined as
\begin{equation}
    T_{\mu \nu} = P g_{\mu \nu} + (P+\mathcal{E}) u_{\mu}u_{\nu},
\end{equation}
where $g_{\mu \nu}$ represents the metric tensor, $P$ is the pressure, $\mathcal{E}$ is the energy density, and $u_{\mu}$ is the four-velocity. For static spherically symmetric stars, the Schwarzschild metric is defined as \cite{1977ApJ...217..799C}
\begin{equation}\label{metric}
    ds^{2} = e^{\nu(r)} dt^{2} -e^{\lambda(r)} dr^{2} - r^{2} (d\theta^{2} + sin^{2} \theta d\phi ^{2}),
\end{equation}
where $e^{\nu(r)}$ and $e^{\lambda(r)}$ are the metric functions. One obtains the Tolman Oppenheimer Volkoff (TOV) equations \cite{PhysRev.55.364, PhysRev.55.374} for the equilibrium structure of NSs by solving the Einstein field equation with the above-defined metric,
\begin{equation}\label{tov1}
\frac{dP(r)}{dr}= -\frac{[\mathcal{E}(r) +P(r)][m(r)+4\pi r^3 P(r)]}{r^2(1-2m(r)/r) } ,
\end{equation}
\begin{equation}\label{tov2}
\frac{dm(r)}{dr}= 4\pi r^2 \mathcal{E}(r) .
\end{equation}
The metric functions become
\begin{equation}
    e^{\lambda(r)} = (1-2m/r)^{-1} ,
\end{equation}
\begin{align}
    \nu(r) &= log(1-2M/R) + 2\int_R^{r} dr' \frac{e^{\lambda(r')}}{r'^{2}} 
     [m(r')+4\pi r'^{3} P(r')] .
    \end{align}
Combining with the given EoS---i.e., $P(\mathcal{E})$ of the matter---, the TOV equations can be solved for the initial conditions $m(r=0)$ = 0 and $P(r=0)$ = $P_c$, where $P_c$ is the central pressure. The radius of the star is determined by requiring that the energy density vanishes at the surface, $P(R)$ = 0, and the mass of the star is then given by $M$ = $m(R)$.

\subsection{Radial oscillations}
\label{radial}

 Considering a spherically symmetric system with only radial motion, the metric shown in Eq.~\eqref{metric} becomes time-dependent and the Einstein field equations can be used to calculate the radial oscillation properties for a static equilibrium structure \cite{1966ApJ...145..505B}.
Considering the radial displacement $\Delta r$ with the pressure perturbation as $\Delta P$, the small perturbation of the equations governing the dimensionless quantities $\xi$ = $\Delta r/r$ and $\eta$ = $\Delta P/P$ are defined as \cite{1977ApJ...217..799C, 1997A&A...325..217G}
 \begin{equation}\label{ksi}
     \xi'(r) = -\frac{1}{r} \Biggl( 3\xi +\frac{\eta}{\gamma}\Biggr) -\frac{P'(r)}{P+\mathcal{E}} \xi(r),
 \end{equation}
 \begin{equation}\label{eta}
 \begin{split}
          \eta'(r) = \xi \Biggl[ \omega^{2} r (1+\mathcal{E}/P) e^{\lambda - \nu } -\frac{4P'(r)}{P} -8\pi (P+\mathcal{E}) re^{\lambda} \\
     +  \frac{r(P'(r))^{2}}{P(P+\mathcal{E})}\Biggr] + \eta \Biggl[ -\frac{\mathcal{E}P'(r)}{P(P+\mathcal{E})} -4\pi (P+\mathcal{E}) re^{\lambda}\Biggr] ,
      \end{split}
 \end{equation}
 where $\omega$ is the frequency oscillation mode and $\gamma$ is the adiabatic relativistic index
 \begin{equation}
     \gamma = \Biggl( 1+\frac{\mathcal{E}}{P}\Biggr) c_s^{2} ,
 \end{equation}
 with $c_s^{2}$ as the speed of sound squared given by
  \begin{equation}\label{cs}
     c_s^{2} = \Biggl(\frac{dP}{d\mathcal{E}}\Biggr)c^{2} .
 \end{equation}

These two coupled differential equations, Eqs.~\eqref{ksi} and \eqref{eta}, are supplemented with two additional boundary conditions: at the center, where $r$ = 0, and at the surface, where $r$ = $R$. The boundary condition at the center requires that
\begin{equation}
    \eta = -3\gamma \xi 
\end{equation}
must be satisfied. The equation Eq.~\eqref{eta} must be finite at the surface and hence
\begin{equation}
    \eta = \xi \Biggl[ -4 +(1-2M/R)^{-1} \Biggl( -\frac{M}{R} -\frac{\omega^{2} R^{3}}{M}\Biggr)\Biggr]
 \end{equation}
 must be satisfied where $M$ and $R$ correspond to the mass and radius of the star, respectively. The frequencies are computed by
 \begin{equation}
\nu = \frac{\omega}{2\pi} = \frac{s \: \omega_0}{2\pi} ~~(kHz),
\end{equation}
where $s$ is a dimensionless number, while $\omega_0 \equiv \sqrt{M/R^3}$.

We use the shooting method analysis, where one starts the integration for a trial value of $\omega^2$ and a given set of initial values that satisfy the boundary condition at the center.  We integrate towards the surface, and the discrete values of $\omega^2$ for which the boundary conditions are satisfied correspond to the eigenfrequencies of the
radial perturbations.

These equations represent the Sturm-Liouville eigenvalue equations for $\omega$. The solutions provide the discrete eigenvalues $\omega_n ^{2}$ and can be ordered as 
 \begin{equation*}
\omega_0 ^{2} < \omega_1 ^{2} <... <\omega_n ^{2}, 
 \end{equation*}
 where $n$ is the number of nodes for a given NS. For a real value of $\omega$, the star will be stable and for an imaginary frequency, it will become unstable. As there is a set of frequencies in the problem studied here, it is enough for one of them to be imaginary to render the star unstable. Also, since the eigenvalues are arranged in above defined manner, it is important to know the fundamental $f$-mode frequency ($n$ = 0) to determine the stability of the star.  

\subsection{Junction Conditions}
\label{junction}

 By analyzing the small radial perturbations of spherically symmetric stars, one can determine the stability of the stellar configuration based on the frequency of the fundamental mode, $\omega_0$. In the mass-central energy density ($M-\mathcal{E}_c$) diagram, the $\partial M/{\partial\mathcal{E}_c} > 0$ condition implies that $\omega_0^2 \ge 0$, which means that a star is stable,  although there could exist branches where
$\partial M/\partial \mathcal{E}_c >$ 0 and the stars are unstable. The opposite holds always true, namely if the $\partial M/{\partial \mathcal{E}_c} < 0$ condition is fulfilled, then $\omega_0^2 < 0$ implying that the star becomes unstable \cite{harrison1965gravitation}.

It is important to keep in mind that phase conversions in the vicinity of the interface can result from small radial perturbations in hybrid stars with sharp discontinuities.
Two limiting cases—slow and fast conversion timescales—have recently been examined in the analysis of this problem \cite{Pereira:2017rmp}.
In this context, slow refers to a conversion timescale $\tau_{conv}$ that is much longer than the perturbed fluid elements' oscillation period $\tau_{osc}$, and rapid refers to a conversion timescale $\tau_{conv} \ge \tau_{osc}$. The phase-splitting surface remains stationary in the case of rapid conversions and oscillates with the same period of perturbations in the case of slow conversions, despite the complex structure of the phase-changing mechanism. This suggests that the nature of the conversion can be minimized to simple junction conditions on the radial fluid displacement $\xi$ and the associated Lagrangian perturbation of the pressure $\Delta P$ at the interface.

For the slow conversion, these two quantities must be continuous across the phase splitting surface \cite{Pereira:2017rmp}:
\begin{align}\label{j1}
    [\xi]_-^+ & = \xi^+ - \xi^- = 0 , \nonumber \\
    [\Delta P]_-^+ & = \Delta P^+ - \Delta P^- = 0 \, 
\end{align}
where $^+$ and $^-$ represent the function values at each side of the hadron-quark interface. 

For the rapid phase transition, since the composition changes instantaneously, the junction conditions become \cite{Pereira:2017rmp}:
\begin{align}\label{j2}
    [\xi]_-^+ & = \Delta P \Bigg[\frac{1}{r_t P_0'}\Bigg]_-^+ , \nonumber \\
    &[\Delta P]_-^+  = 0 \, 
\end{align}

with $r_t$ being the radial position and $P_0' = dP_0/{dr}$ as the gradient of the pressure at the interface.

A great deal of recent works have investigated the dynamic stability of hybrid stars with rapid and slow interface conversions using the junction conditions of Equations (\ref{j1}) and (\ref{j2}) \cite{Lugones:2023xeq, Pereira:2020cmv, VasquezFlores:2010eq, VasquezFlores:2012vf, PhysRevD.101.103003, Zhang:2023zth}. 

Numerical calculations have shown that even if $\partial M/{\partial \mathcal{E}_c} < 0$, {\ish{$\omega_0^2$}} can be a real number, thus implying stability in hybrid stars with sharp density discontinuities and slow interface conversions. Vasquez Flores \textit{et al.} \cite{VasquezFlores:2012vf} was the first to report this, and Pereira \textit{et al.} \cite{Pereira:2017rmp} was the first to establish the link between these stable configurations and slow interface conversions. Different facets of these novel configurations for particular EOSs have been investigated in later studies \cite{PhysRevD.107.103042, Mariani:2022xek, Rodriguez:2020fhf}.
These hybrid stars where $\partial M/{\partial \mathcal{E}_c} < 0$ are referred to as slow-stable hybrid stars (SSHSs) \cite{Lugones:2023xeq}. 
This study extends the previous works with slow phase conversion by including a phase deconfinement transition to quark matter and emphasizing the interactions between hyperons and $\Delta$ baryons as well.

\section{Numerical results and discussion}
\label{results}

 We shall now present and discuss our main results regarding the properties of the compact objects studied here, namely EoSs and mass-to-radius relationships, speed of sound and adiabatic relativistic index, dimensionless deformability, and frequencies of radial oscillation modes. In the analysis performed in this work, there is an interplay between four different matter compositions on the one hand, and on the other hand a slow phase transition between a quark core and a nucleonic mantle. The impact of the different matter compositions on the properties of the stars has been discussed in detail in \cite{PhysRevD.107.123022}. Therefore, in the discussion to follow we shall focus ourselves on the new effects due to the presence of the quark core.

\subsection{EoS and M-R Profile}
\label{mr}

\begin{figure}[ht]
\centering
	\includegraphics[scale=0.35]{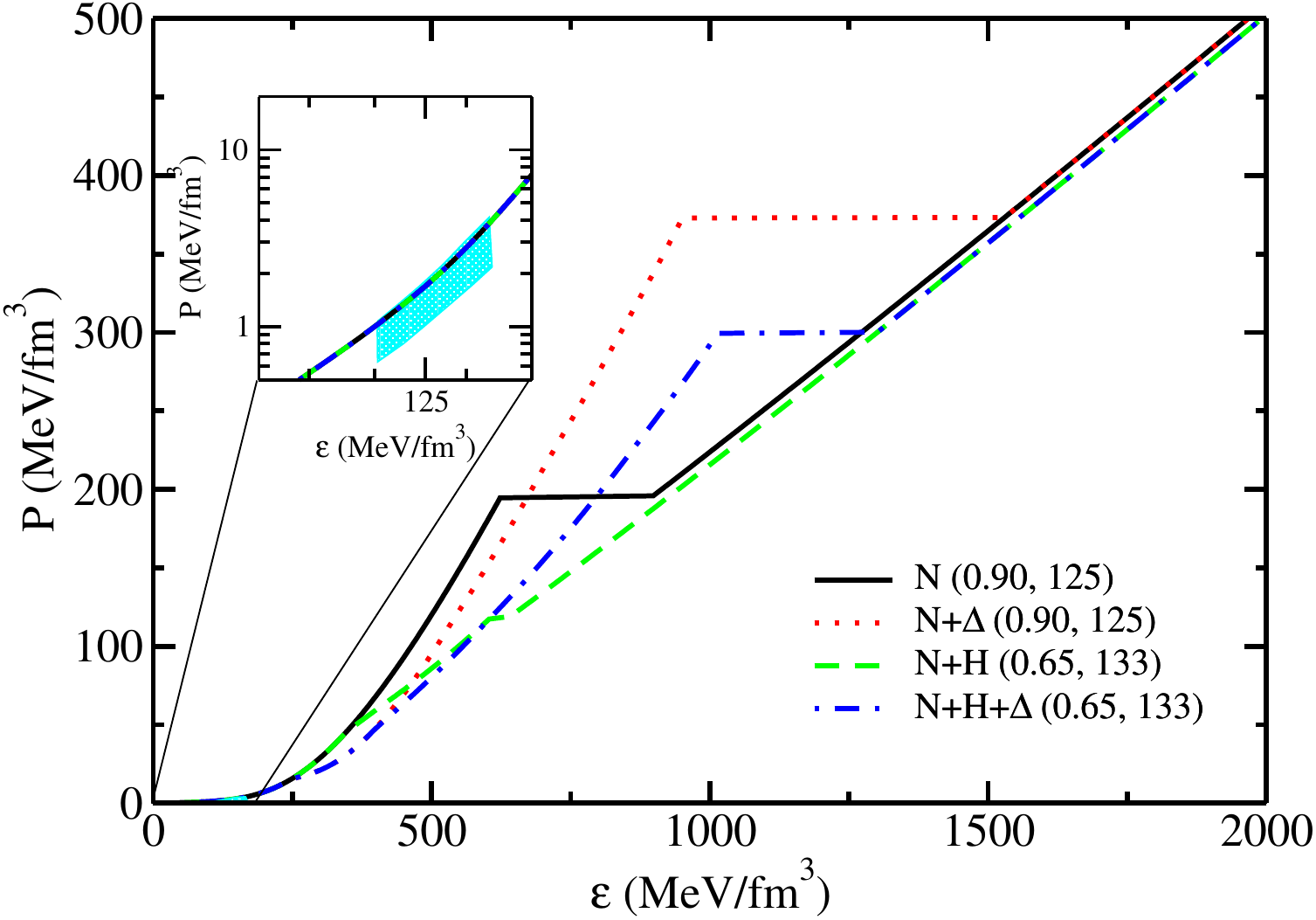}
	\caption{Energy density and pressure variation for the N (solid lines), N+${\Delta}$ (dotted lines), N+H (dashed lines), and N+H+${\Delta}$ (dot-dashed lines) DD-RMF EoS for ${\alpha_v}$ = 1.0 with a phase transition to the quark matter at parameter values (${C, D^{1/2}}$) = (0.90, 125 MeV) for N and N+${\Delta}$ EoS and  (${C, D^{1/2}}$) = (0.65, 133 MeV) for N+H and N+H+$\Delta$ EoS. The inset represents the EoS at low density with constraints from chiral effective field theory \cite{Hebeler:2013nza}.}
	\label{fig1} 
\end{figure}

Figure \ref{fig1} shows the EoS---variation of pressure with energy density---for an NS in beta-equilibrium and charge-neutral conditions in the standard units of nuclear physics, for four different hadronic matter compositions and two different pairs of the $C, D$ parameters. Contrary to NSs made of pure nucleonic matter plus exotic particles without the quark core, here we observe a few distinctive features, which may be summarized as follows: At intermediate energy densities specifically at the coexistence pressure, EoSs reach a plateau and then they increase again after the phase transition. The pure nucleonic EoS is not the stiffest anymore throughout the whole energy density range. Rather, the $N$+$\Delta$ EoS becomes the stiffest at intermediate values of the energy density, since the coexistence point is at a higher pressure.

In the region of low to intermediate density, before the phase transition occurs, stiff EoS is produced by pure nucleonic matter. When $\Delta$s are added to the nuclear matter, the EoS becomes softer. This is because the inclusion of new particles results in a distribution of the Fermi pressure among the many particles with more degrees of freedom, softening the EoS. When only nucleons are considered in neutron star matter, the effective nucleon mass, $M_n$, decreases gradually with increasing $n_B$. However, adding other baryon species like hyperons or $\Delta$s leads to a much faster decrease in the nucleon effective mass. This results from each additional particle contributing negatively to $M_n$,  due to the addition of an extra term per particle kind via the scalar density dependence of the $\sigma$ field, causing it to drop more rapidly as the baryon multiplicity increases. For some cases, $M_n$ reaches zero at a density lower than the maximum $n_B$ allowed in the numerical calculations (usually around 1.5 fm$^{-3}$). While this behaviour was previously well-documented in hypernuclear star matter \cite{Schaffner:1995th}, the inclusion of $\Delta$ particles exacerbates the effect, making the rapid decline even more pronounced. A detailed explanation of how this behaviour affects hadronic matter with $\Delta$ baryons is given in Ref. \cite{PhysRevC.106.055801}, where this lower maximum EoS density is reflected astrophysically in having Mass-Radius profile curves that do not reach the maximum mass stability point. Here, however, the phase transition point always occurs at densities lower than the vanishing of $M_n$, which prevents this unwanted behaviour from manifesting.

The addition of ${\Delta}$s in the hyperonic matter, N+H+${\Delta}$, is more intricate, even though the hyperons further soften the EoS. As can be observed in Figure \ref{fig1}, the N+H+${\Delta}$ is softer than the N+H composition at lower densities. The EoS with the N+H+${\Delta}$ composition becomes stiffer than the N+H composition as the density increases. This can be attributed to the fact that a neutron-electron pair, which is preferred over light baryons due to the attractive potential, is replaced at the top of their Fermi seas by the ${\Delta^{-}}$ baryon. Later on, we observe the electric charge-neutral particles ${\Lambda^{0}}$ and ${\Delta^{0}}$. The outer crust section of the non-unified EoS is modeled using the Baym-Pethick-Sutherland (BPS) EoS \cite{Baym:1971pw}. The DD-RMF parameter set in the Thomas-Fermi approximation is employed to produce the EoS in the non-uniform matter for the inner crust  \cite{PhysRevC.79.035804, PhysRevC.94.015808,rather2020effect}. The EoS at the low-density region satisfies the constraint obtained from chiral effective field theory \cite{Hebeler:2013nza}, as shown in the inset of Figure \ref{fig1}.

 As for the phase transition, we observe that the presence of ${\Delta}$s shifts the coexistence point towards higher densities for the same deconfined EoS, relating to the above-mentioned effect. The EoS at higher densities---after the phase transition---is a lot more uniform than the hadronic counterpart, as we observe that the parameter set of (${C, D^{1/2}}$) = (0.90, 125 MeV) only produces a slightly harder EoS than the one of (${C, D^{1/2}}$) = (0.65, 133 MeV). On the other hand, the most significant influence comes from where the coexistence point is when building the hybrid EoS.

\begin{figure*}[htbp!]
\centering
    \includegraphics[width=0.47\linewidth]{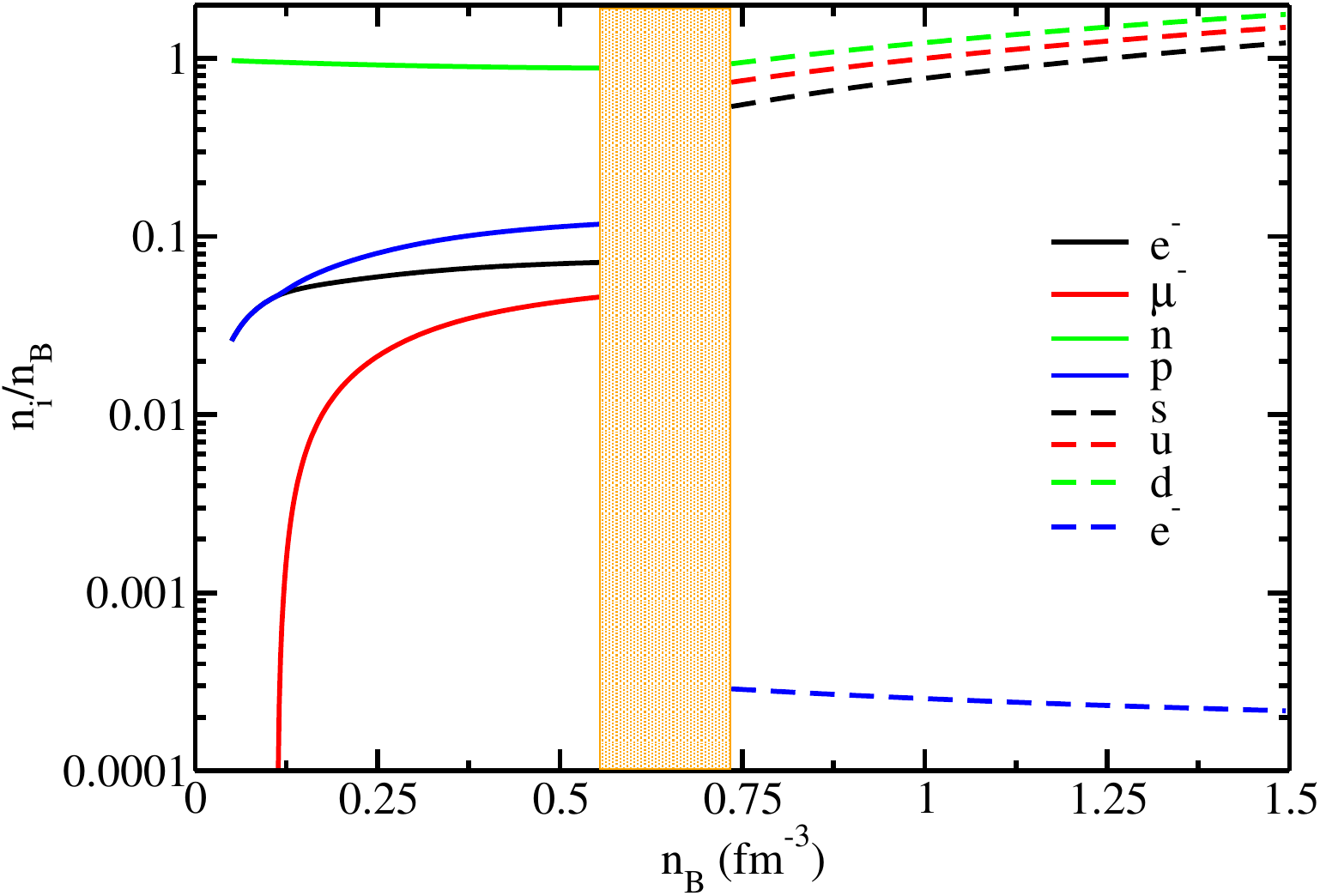}\hfil
.    \includegraphics[width=0.47\linewidth]{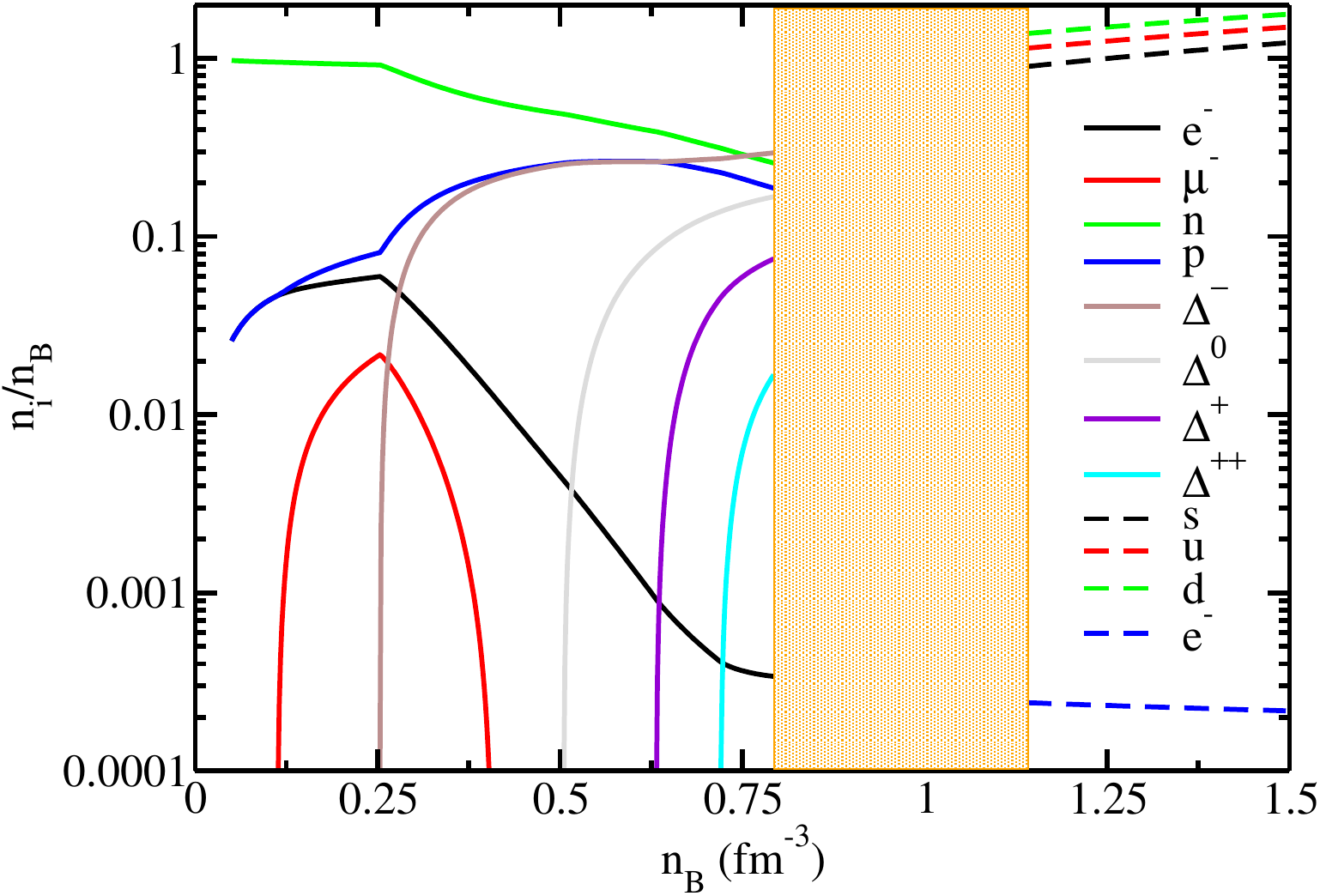}\par\medskip
    \includegraphics[width=0.47\linewidth]{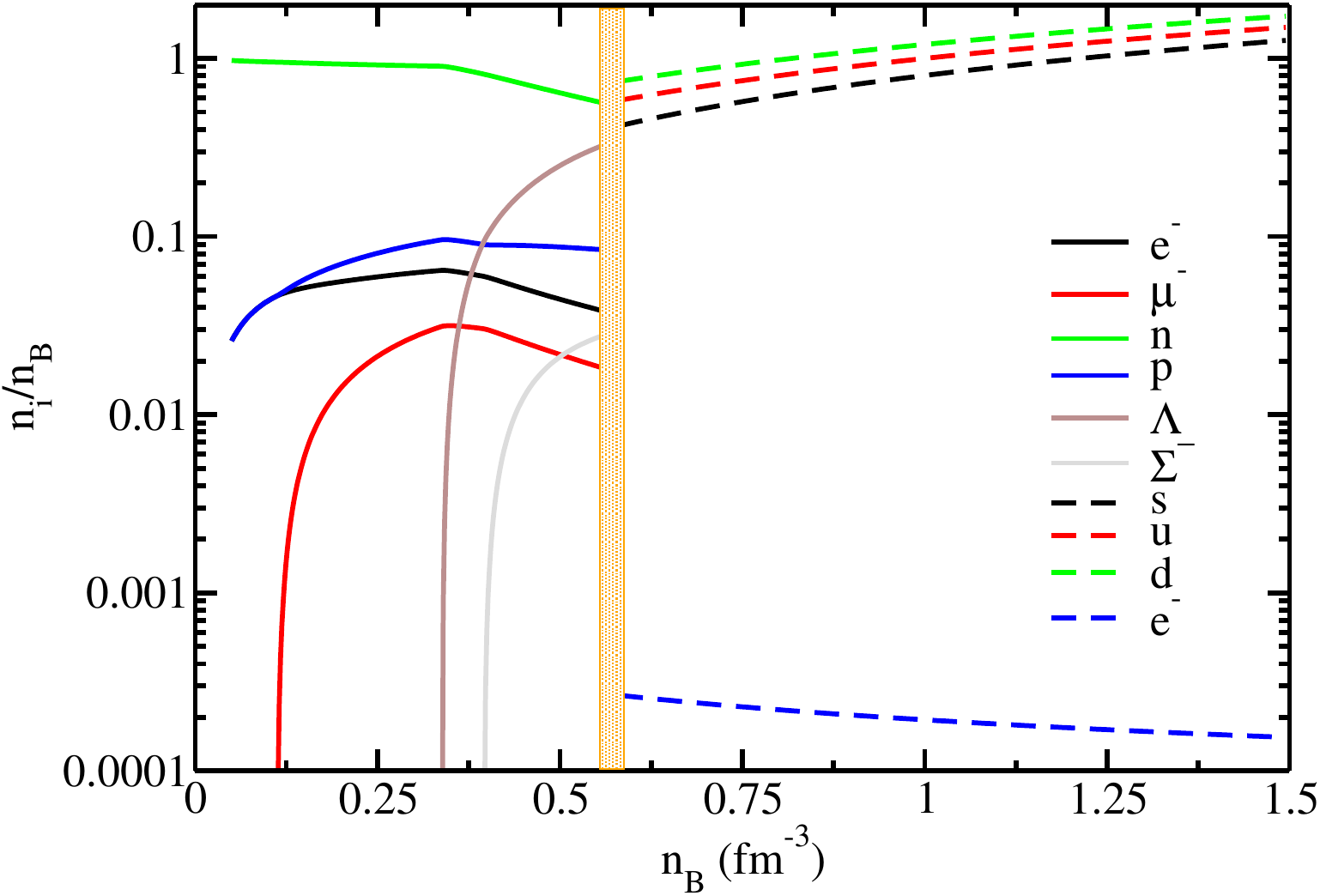}\hfil
    \includegraphics[width=0.47\linewidth]{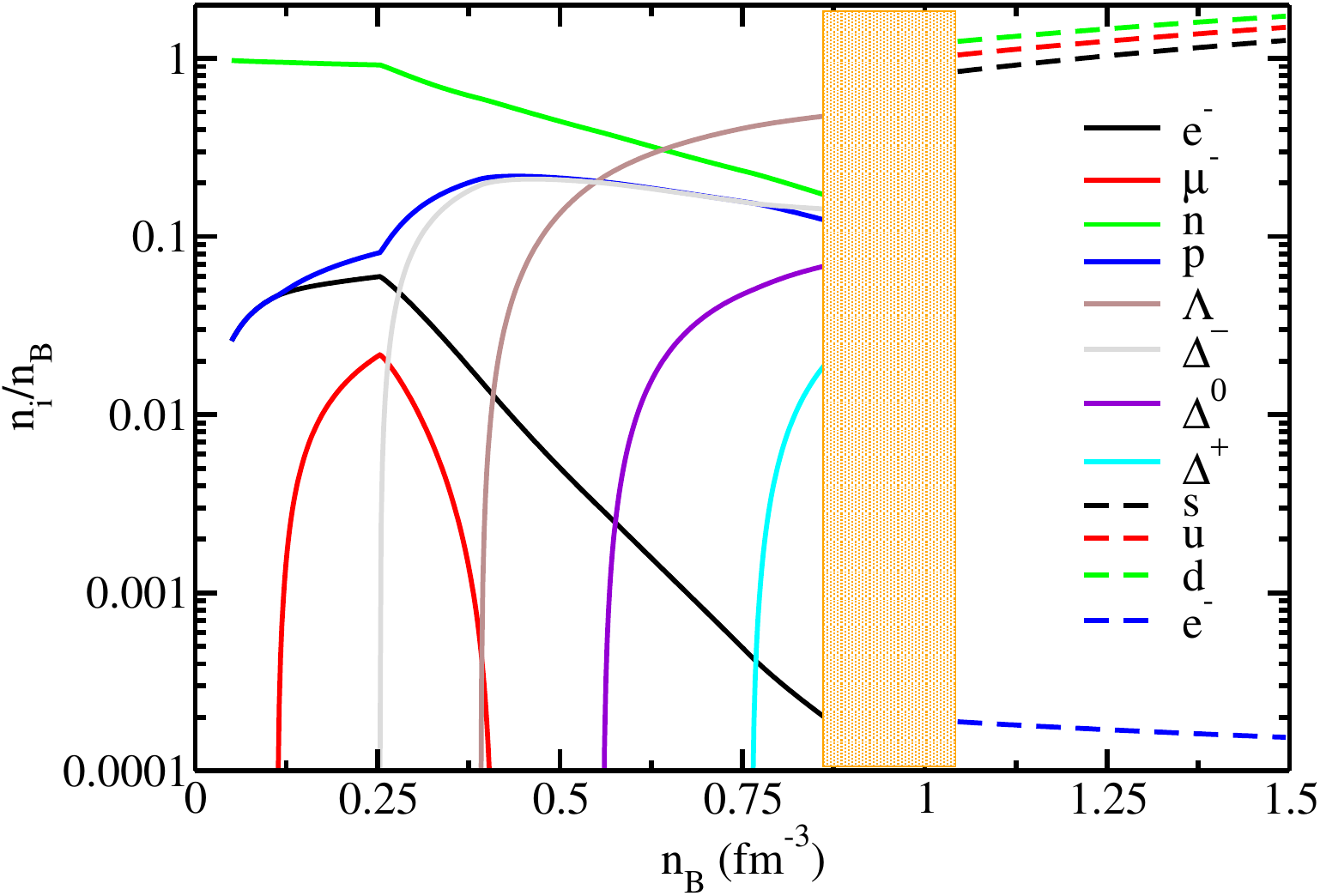}
\caption{Particle population for different hybrid EoS: N (top left), N+${\Delta}$ (top right), N+H (bottom left), N+H+${\Delta}$ (bottom right). The solid lines represent baryons while the dashed lines represent the quark fractions. The shaded region (orange) represents the density discontinuity gap between both parts of the EoS.}
\label{fig:pp}
\end{figure*}

  Figure \ref{fig:pp} shows the particle population plots for the hybrid EoSs considered in this study. We consider that the hadron-quark deconfinement is a first-order phase transition, so it imposes a density discontinuity gap between both components. Due to that, e.g., the pure N hadronic composition shows a density jump of 0.183 fm$^{-3}$ between the phases. As discussed earlier, the inclusion of $\Delta$ baryons shifts the coexistence point to higher densities, resulting in a mixed phase region of 0.348 fm$^{-3}$ for the N+$\Delta$ composition. Similarly, the N+H hybrid EoS exhibits a very narrow mixed-phase region, that broadens and moves to higher $n_B$ when $\Delta$ baryons are added to the composition (N+H+$\Delta$). The transition density from hadronic matter to the quark matter for each EoS is shown in Table \ref{tab:coex}.

  \begin{figure}[htbp!]
\centering
  	\includegraphics[scale=0.35]{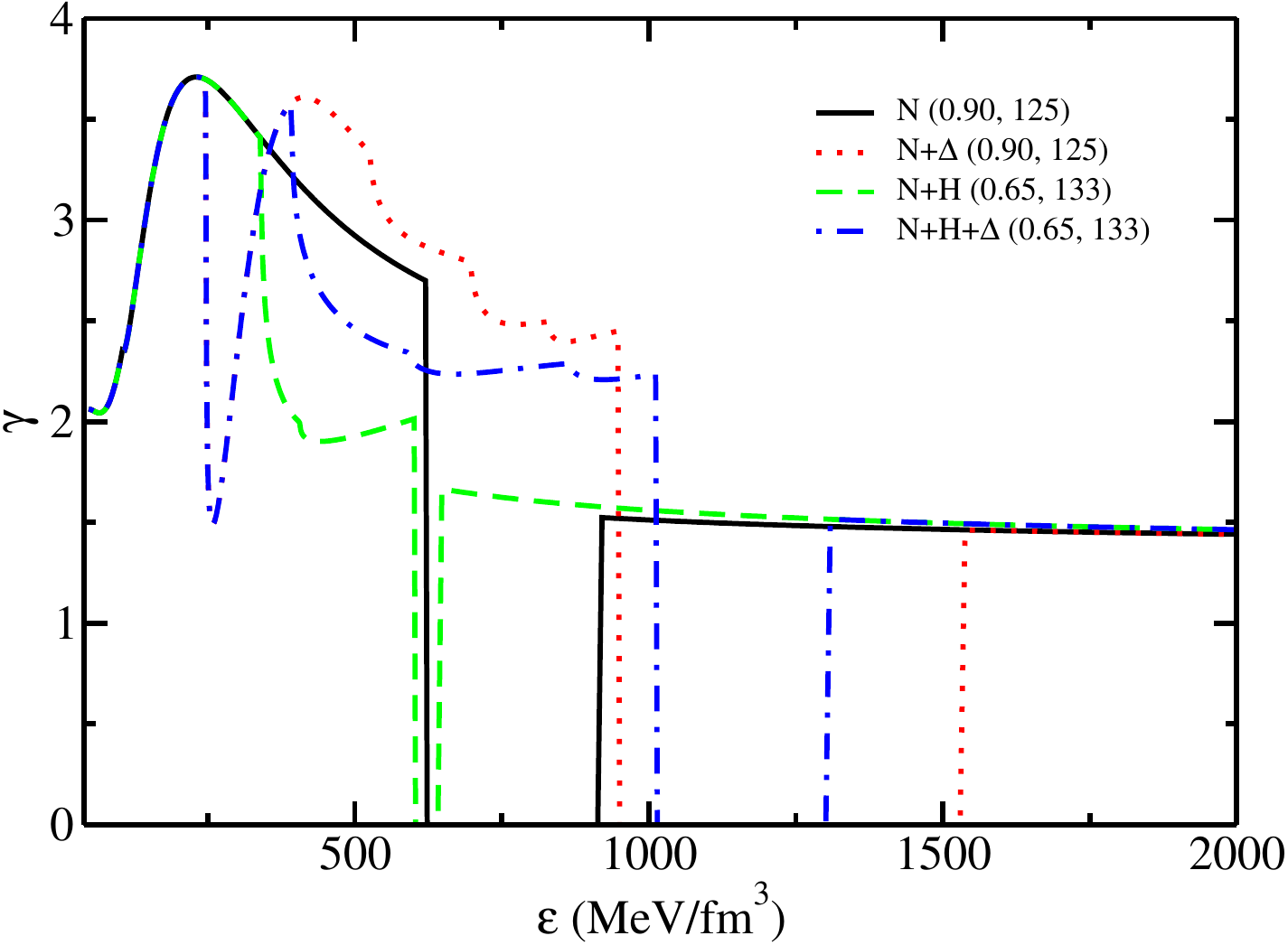}
  	\caption{Adiabatic index as a function of energy density for the hybrid stars with N (solid lines), N+${\Delta}$ (dotted lines), N+H (dashed lines), and N+H+${\Delta}$ (dot-dashed lines) composition at parameter values (${C, D^{1/2}}$) = (0.90, 125 MeV) for N and N+${\Delta}$ EoS and  (${C, D^{1/2}}$) = (0.65, 133 MeV) for N+H and N+H+${\Delta}$ EoS. }
  	\label{fig2a} 
  \end{figure}

Figure \ref{fig2a} displays the adiabatic index ${\gamma}$ as a function of energy density for hybrid EoS with different matter compositions. For hybrid EoS with N only, the ${\gamma}$ increases to a peak value of around 3.7 at low energy density and then drops smoothly at around 600 MeV/fm$^3$ energy density before the phase transition to the deconfined quark matter begins, causing the adiabatic index goes to zero. The adiabatic index for the pure quark matter after the phase transition remains almost unchanged with a value of 1.5. For the hybrid EoS with N+$\Delta$, since the phase transition occurs at high density, the adiabatic index attains a larger value than hybrid N EoS in the energy density range of 400-900 MeV/fm$^3$. For hybrid EoS with N+H, the EoS is softened by the presence of hyperons, particularly ${\Lambda^0}$, and $\gamma$ decreases to around ${\approx}$ 350 MeV/fm$^3$. The onset of ${\Delta^-}$ baryons allows the value of ${\gamma}$ for hybrid N+H matter containing ${\Delta}$ baryons to drop rapidly at about ${\approx}$ 250 MeV/fm$^3$. However, the ${\gamma}$ also grows with density and surpasses the hybrid N EoS. It is not seen that the hyperonic matter exhibits this significant rise in ${\gamma}$ behavior. We see a sharp decline in the value ${\gamma}$ for N+H+${\Delta}$ matter because of the ${\Delta^-}$ threshold, which is followed by a brief increase and a further decrease because of the onset of ${\Lambda^0}$ hyperon. For all the hybrid EoS, the $\gamma$ value remains almost the same after the phase transition.
  
  \begin{figure}[h]
  \centering
  	\includegraphics[scale=0.35]{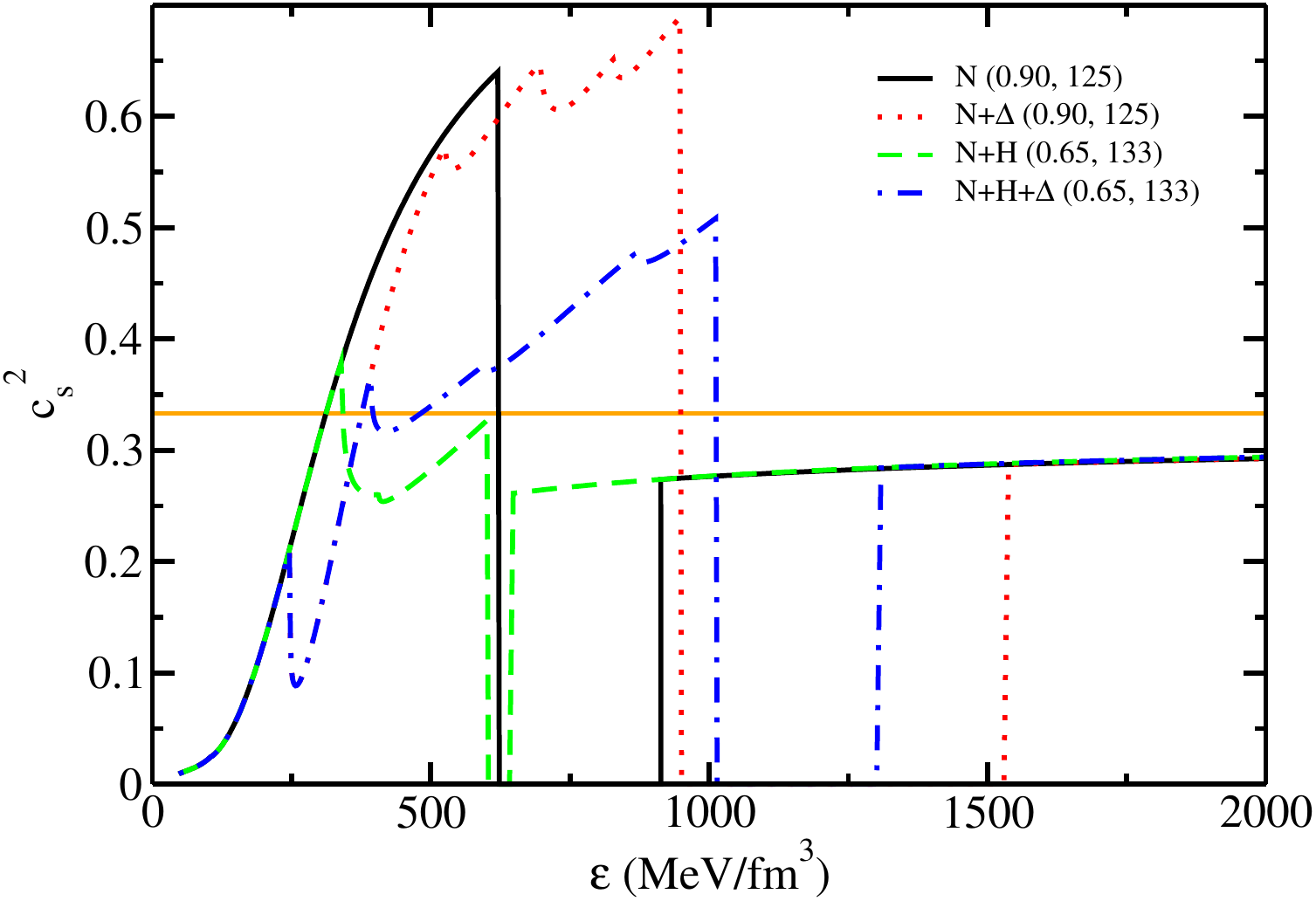}
  	\caption{Speed of sound squared as a function of energy density for the hybrid stars with N (solid lines), N+${\Delta}$ (dotted lines), N+H (dashed lines), and N+H+${\Delta}$ (dot-dashed lines) composition at parameter values (${C, D^{1/2}}$) = (0.90, 125 MeV) for N and N+ ${\Delta}$ EoS and  (${C, D^{1/2}}$) = (0.65, 133 MeV) for N+H and N+H+${\Delta}$ EoS. The solid orange line represents the conformal limit ${c_s^{2}}$ = 1/3.}
  	\label{fig2} 
  \end{figure}
  
The behavior of the squared speed of sound as a function of energy density for hybrid EoS with various compositions of the matter under consideration in this work is shown in Figure \ref{fig2}. It is a significant quantity that conveys information regarding gravitational wave signatures, tidal deformability, and shear viscosity \cite{PhysRevC.102.055801, Lopes_2021}. 


Thermodynamic stability demands that ${c_s^{2} > 0}$, whereas causality sets an upper bound of ${c_s^{2}}$ ${\leq 1}$ for the permitted range of the sound speed squared. Perturbative QCD results predict an upper limit of ${c_s^{2}}$ = 1/3 for very high densities \cite{PhysRevLett.114.031103}. Several investigations \cite{PhysRevLett.114.031103, PhysRevC.95.045801, Tews_2018} have reported that the two solar mass requirements need a speed of sound squared greater than the conformal limit (${c_s^{2}}$ = 1/3), indicating that the matter inside NS is a highly interacting system.

From Figure \ref{fig2}, the hybrid EoS with N only presents a very large value of the ${c_s^{2}}$ for the pure nucleonic matter. The phase transition to the quark matter lowers the value to zero at the coexistence point and then further increases for the pure quark matter. The value of the ${c_s^{2}}$ for the pure quark matter is lower than that for the pure nucleonic matter. For hybrid EoS with different particle compositions, one can see the kinks present at the onset of various particles before the phase transition. At lower densities, the pure baryonic part of ${c_s^{2}}$ for hybrid N, N+$\Delta$, and N+H+$\Delta$ EoS violate the conformal limit. The curve for the hybrid N+H+${\Delta}$ composition predicts a higher value of the speed of sound squared at intermediate densities because of the early appearance of ${\Delta^{-}}$ particles and a delayed phase transition to the quark matter. 
Finally, at high energy densities all four speeds of sounds remain below the conformal limit, contrary to what was observed in \cite{PhysRevD.107.123022}, where at high energy densities two of them remained above and the other two below the conformal limit. This is expected since we now have a deconfined EoS at high energy densities---which yields a speed of sound that should approach the conformal limit from below.

  \begin{figure}
  \centering
 	\includegraphics[scale=0.35]{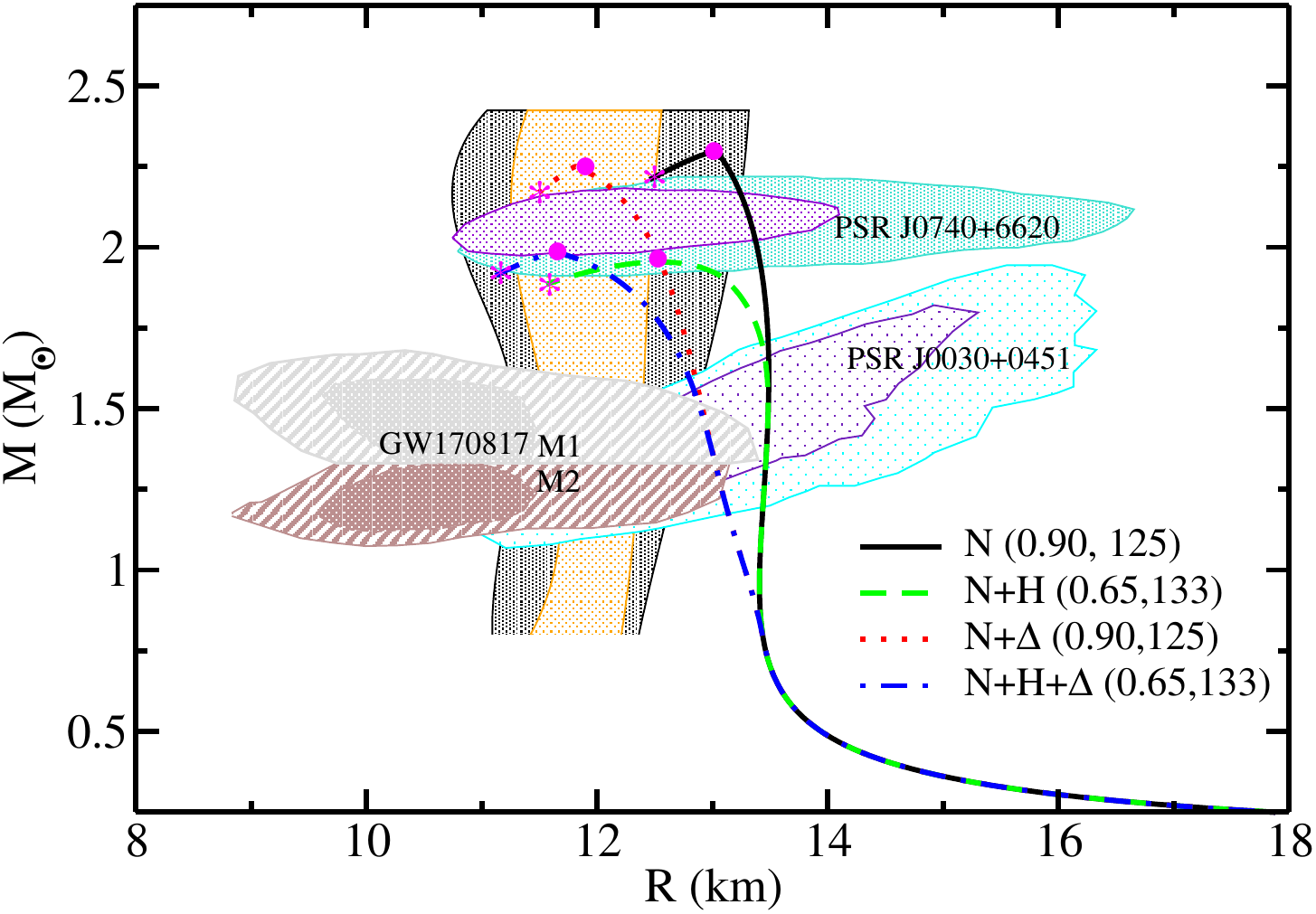}
 	\caption{Mass-Radius profile for DD-RMF parameter set with different compositions of $\Delta$ baryons and hyperons along with a phase transition to the quark matter at parameter values (${C, D^{1/2}}$) = (0.90, 125 MeV) for N and N+$\Delta$ EoS and (${C, D^{1/2}}$) = (0.65, 133 MeV) for N+H and N+H+${\Delta}$ EoS. The 68\% (violet) and 95\% (turquoise) credible regions for mass and radius are inferred from the analysis of PSR J0740+6620 \cite{miller2021,2021ApJ...918L..27R}. For PSR J0030+0451, the indigo dotted region is for 68\% credibility while the green dotted region is for 95\% credibility \cite{Miller_2019a}. The grey upper (brown lower) shaded region corresponds to the higher (smaller) component of the GW170817 event \cite{PhysRevLett.121.161101}. The joint constraints from HIC experiments and multimessenger astrophysics with 68\% (orange) and 95\% (black) credible ranges are also shown \cite{Huth:2021bsp}. }
 	\label{fig3} 
 \end{figure}

 Solving the TOV equations for different EoSs, we display the mass-radius profiles for hybrid EoSs with different matter compositions in Figure \ref{fig3}. For hybrid EoS with N only, a maximum mass of 2.30 ${M_{\odot}}$ with a radius of 13.03 km is achieved. The appearance of $\Delta$ baryons softens the EoS and hence the maximum mass and the corresponding radius decrease to a value of 2.25 ${M_{\odot}}$ and 11.80 km, respectively. The radius at the canonical mass for N and N+$\Delta$ is 13.47 and 12.98 km, respectively. For hybrid EoS with N+H, the maximum mass obtained is 1.97 ${M_{\odot}}$ with a radius of 12.47 km. With the addition of $\Delta$s, the EoS stiffens and the maximum mass increases to a value of 1.98 ${M_{\odot}}$  with an 11.57 km radius. 
 The MR profile for the hybrid EoS with N and N+$\Delta$ satisfy the 2.0 ${M_{\odot}}$ limit as well as the mass and the radius constraints from various measurements \cite{miller2021,2021ApJ...918L..27R, Miller_2019a} excluding the GW170817 constraint \cite{PhysRevLett.121.161101}. As for the N+H and N+H+${\Delta}$ hybrid EoS profiles, although they satisfy the lower limit of the PSR J0740+6620 for mass and radius \cite{miller2021,2021ApJ...918L..27R}, they fail to produce a 2.0 ${M_{\odot}}$ hybrid star. 
 
Even though the quark parameter values, $C$ and $D^{1/2}$ were chosen to obtain a stiff quark EoS, the appearance of hyperons and then a deconfined phase transition to the quark matter softens the EoS enough to produce a star with a maximum mass less than 2 ${M_{\odot}}$. All four EoSs satisfy the joint constraints from HIC experiments and multimessenger astrophysics with 68\% (orange) and 95\% (black) credible ranges at their respective maximum masses \cite{Huth:2021bsp}. 
 We also observe that with the additional degree of freedom ($\Delta$s), the MR profiles satisfy this constraint beyond 1.5 $M_{\odot}$.

The transition density in all our hybrid models, especially for EoSs with $\Delta$s, is high and since we have used a slow phase conversion in this work and applied the necessary junction conditions at the phase transition, the configurations that occur to the left of the maximum mass star are also stable. In other words, their central densities are greater than the central density of the star with $M_{max}$. 
  In Figure \ref{fig3}, the magenta dot on each MR curve represents the maximum mass configuration obtained, while the star on each curve represents the point where the $f$-mode
frequency tends to zero, occurring beyond the maximum mass configuration. The region between the dot mark and the star corresponds to the slow stable hybrid stars (SSHSs). As pointed out in Ref. \cite{Lugones:2021bkm}, the length of the SSHS branch depends upon the energy density jump between two phases and the stiffness of the quark EoS. The smaller the energy density jump between the two phases, the larger the length of the SSHS. For N+H hybrid EoS, the density jump $\Delta \mathcal{E}$ between the hadronic and quark phase is around 38 MeV/fm$^{3}$, the smallest among all hybrid EoSs. Also, since for the N+H Hybrid EoS a smaller value of $C$ parameter for the quark EoS is employed---which provides a stiff EoS---the length of the SSHS is large with a radius of around 0.88 km.

\begin{figure}[htp!]
\centering
	\includegraphics[scale=0.35]{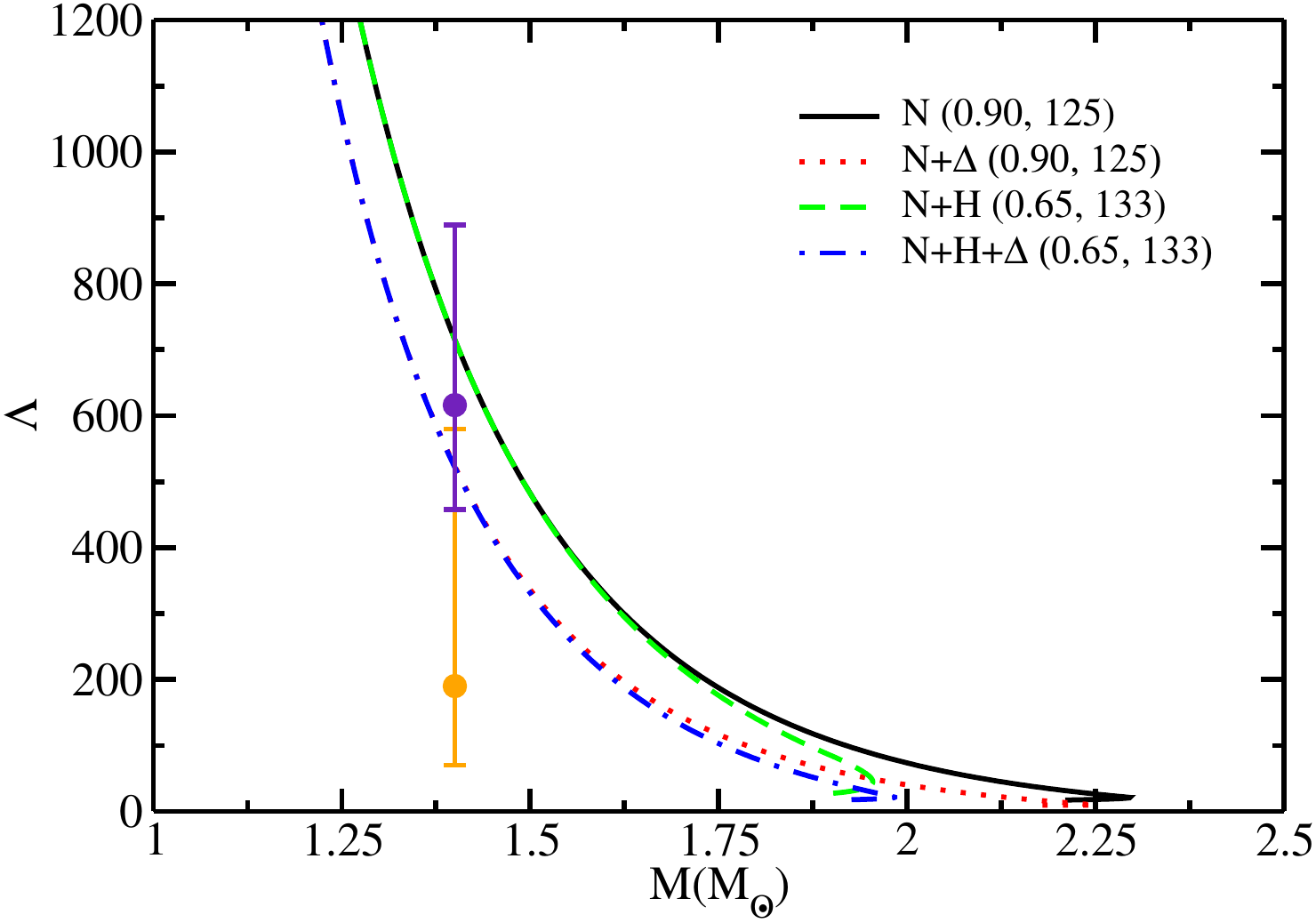}
	\caption{ Dimensionless tidal deformability variation with the mass for different compositions of ${\Delta}$ baryons and hyperons along with a phase transition to the quark matter at parameter values  (${C, D^{1/2}}$) = (0.90, 125 MeV) for N and N+${\Delta}$ EoS and (${C, D^{1/2}}$) = (0.65, 133 MeV) for N+H and N+H+${\Delta}$ EoS. The orange and violet error bars represent the ${\Lambda}$ = 190${^{+390}_{-120}}$ constraints concluded by the LVC measurement from GW170817 \cite{PhysRevLett.121.161101} and $\Lambda$ = 616${^{+273}_{-158}}$ constraint based on the assumption that GW190814 is a NS-BH binary \cite{Abbott:2020khf}, respectively. }
	\label{fig:tidal} 
\end{figure}

Figure \ref{fig:tidal} displays the dimensionless tidal deformability, $\Lambda$, as a function of the stellar mass, $M$, for Hybrid EoSs for the four different matter compositions considered here. As usual, $\Lambda$ is large for low stellar masses and small for high stellar masses, decreasing monotonically with $M$.
For the hybrid EoS with N and N+H, the dimensionless tidal deformability at the canonical mass, i.e. at $1.4~M_{\odot}$, is computed to be $\Lambda=720.50$, satisfying the $\Lambda$ = 616${^{+273}_{-158}}$ constraint based on the assumption that GW190814 is a NS-BH binary \cite{Abbott:2020khf}. For hybrid EoS with N+$\Delta$ and N+H+$\Delta$, the softness of the EoS predicts a lower value of dimensionless tidal deformability, $\Lambda_{1.4 M_{\odot}}$ = 520.36, which satisfies the constraint of ${\Lambda}$ = 190${^{+390}_{-120}}$ by the LVC measurement from GW170817 \cite{PhysRevLett.121.161101}.
Table \ref{tablemr}  displays the stellar properties of the different hybrid EoSs. The quantities with subscript $max$ represent the maximum mass configurations, while the ones with subscript $*$ represent the stable branch configurations beyond the maximum mass.

\begin{table}[htb!]
\centering
		\caption{Stellar star properties for N and N+${\Delta}$ EoS at quark parameter values (${C, D^{1/2}}$) = (0.90, 125 MeV), and N+H and N+H+${\Delta}$ EoS at quark parameter values (${C, D^{1/2}}$) = (0.65, 133 MeV). Mass is in ($M_{\odot}$), radius in (km), and central energy density in (MeV. fm$^{-3}$). \label{tablemr} }
\begin{tabular}{ p{1.3cm}p{1.3cm}p{1.3cm}p{1.3cm}p{1.3cm}p{1.3cm}p{1.3cm}p{1.3cm}p{1.3cm}}
 \hline
 EoS & $M_{max}$&$R_{max}$& $\mathcal{E}_{c, max}$ & $M_{*}$&$R_{*}$  &$\mathcal{E}_{c, *}$ & $R_{1.4 M_{\odot}}$ & $\Lambda_{1.4 M_{\odot}}$\\
 \hline
N    &2.30  &13.03&917 & 2.22&12.50&1392  &13.47&720.50\\
N+$\Delta$ &2.25	&11.80&1540&2.17&11.49&2172	&12.98	&520.36\\
N+H &1.97&12.47&984	&1.89&11.59&1674&13.47	&720.50\\
N+H+$\Delta$ &1.98	&11.57&1408&1.92&11.16&1948	&12.97	&520.36\\

\hline 
\end{tabular}
\end{table}

\subsection{Radial Profiles}
\label{profile}

\begin{figure*}[htbp!]
\centering
    \includegraphics[width=0.47\linewidth]{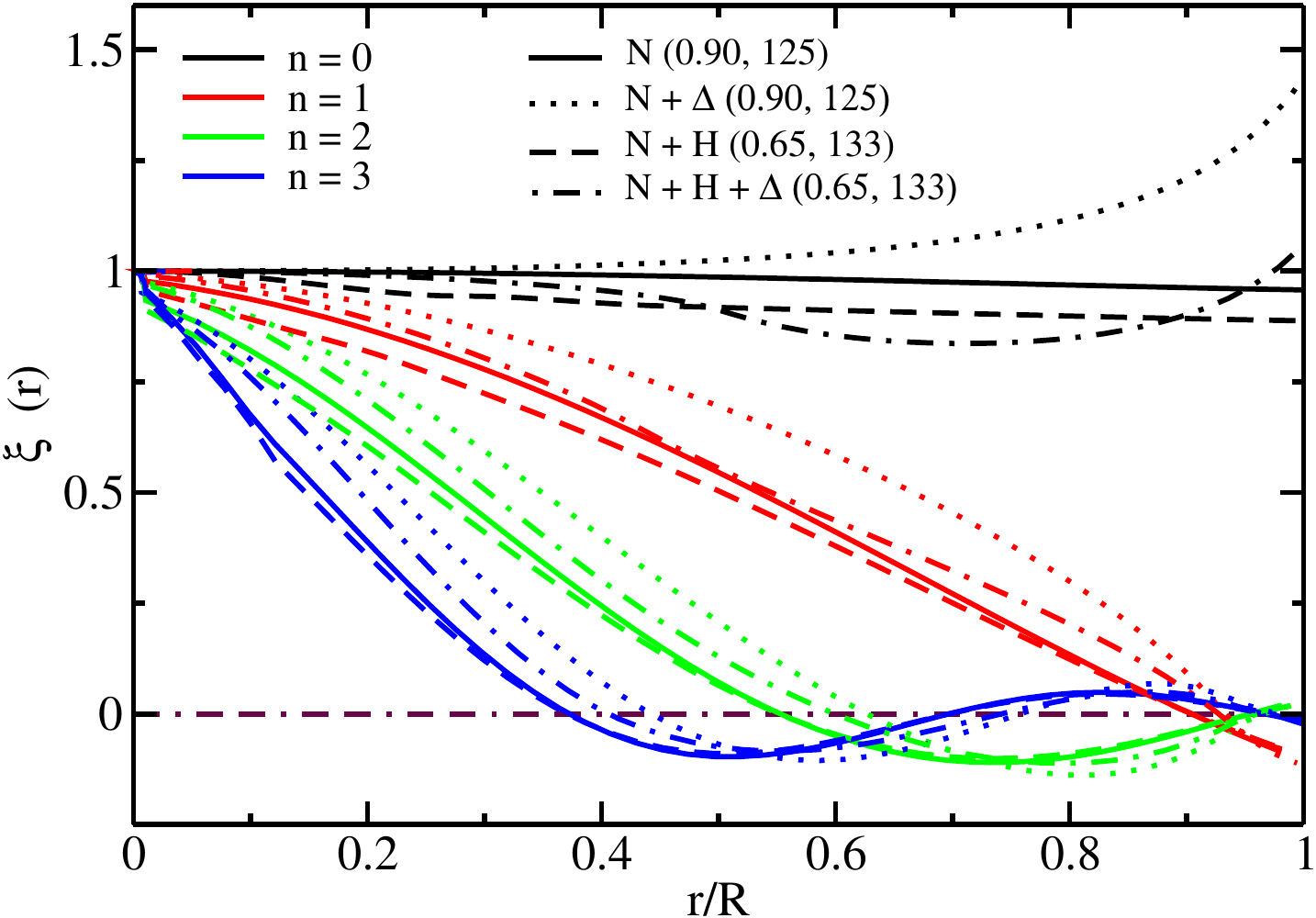}\hfil
.    \includegraphics[width=0.47\linewidth]{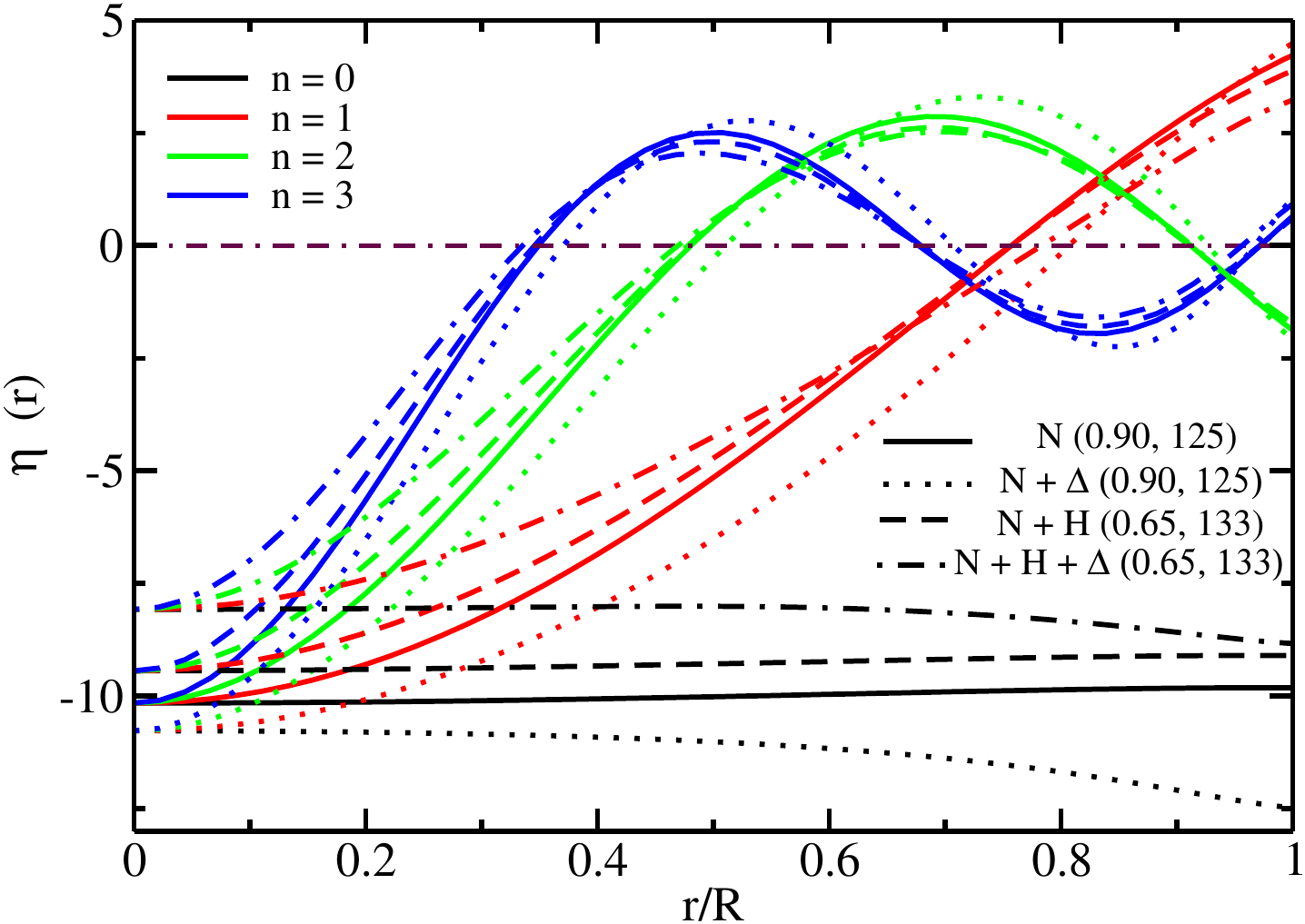}\par\medskip
    \includegraphics[width=0.47\linewidth]{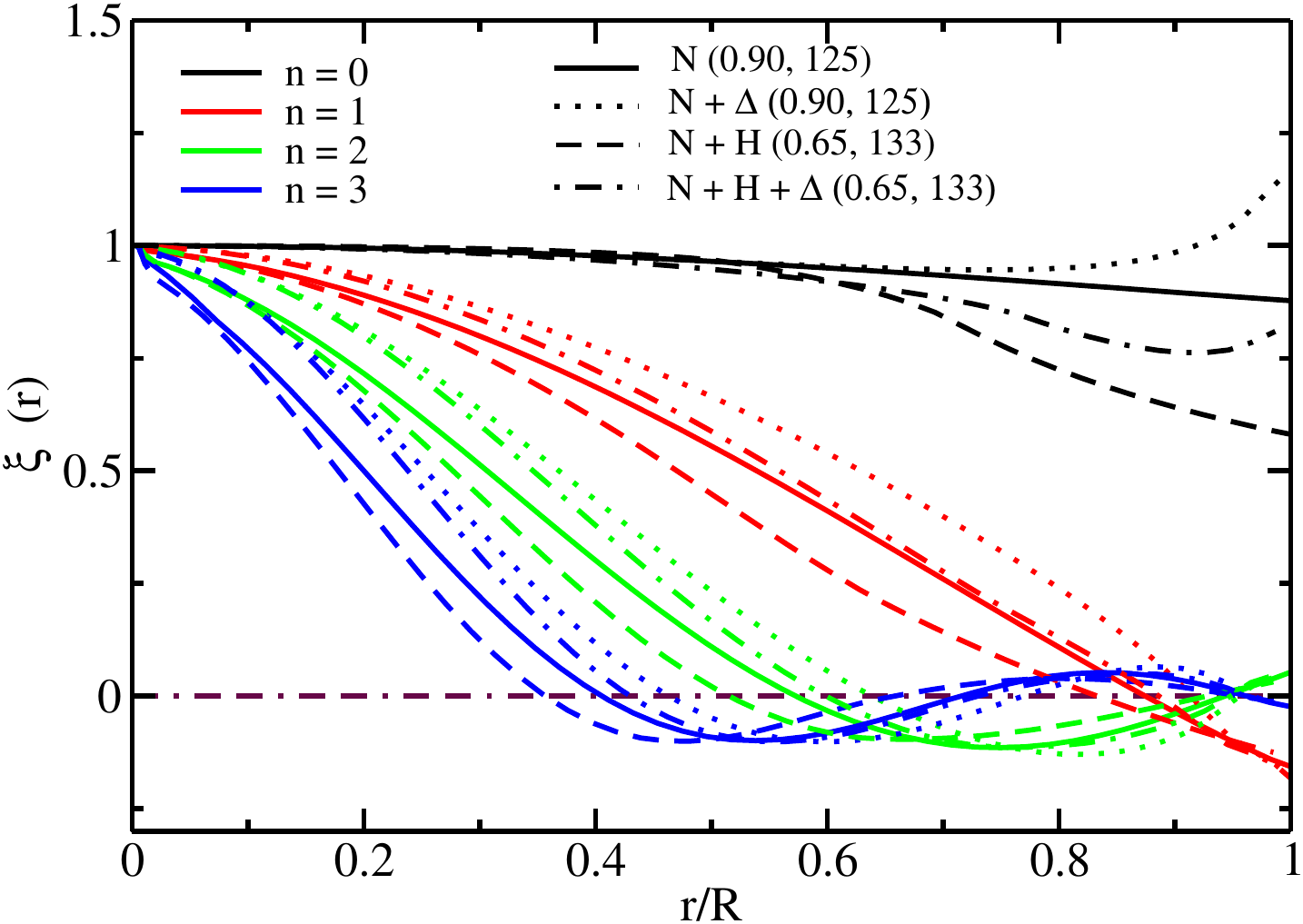}\hfil
    \includegraphics[width=0.47\linewidth]{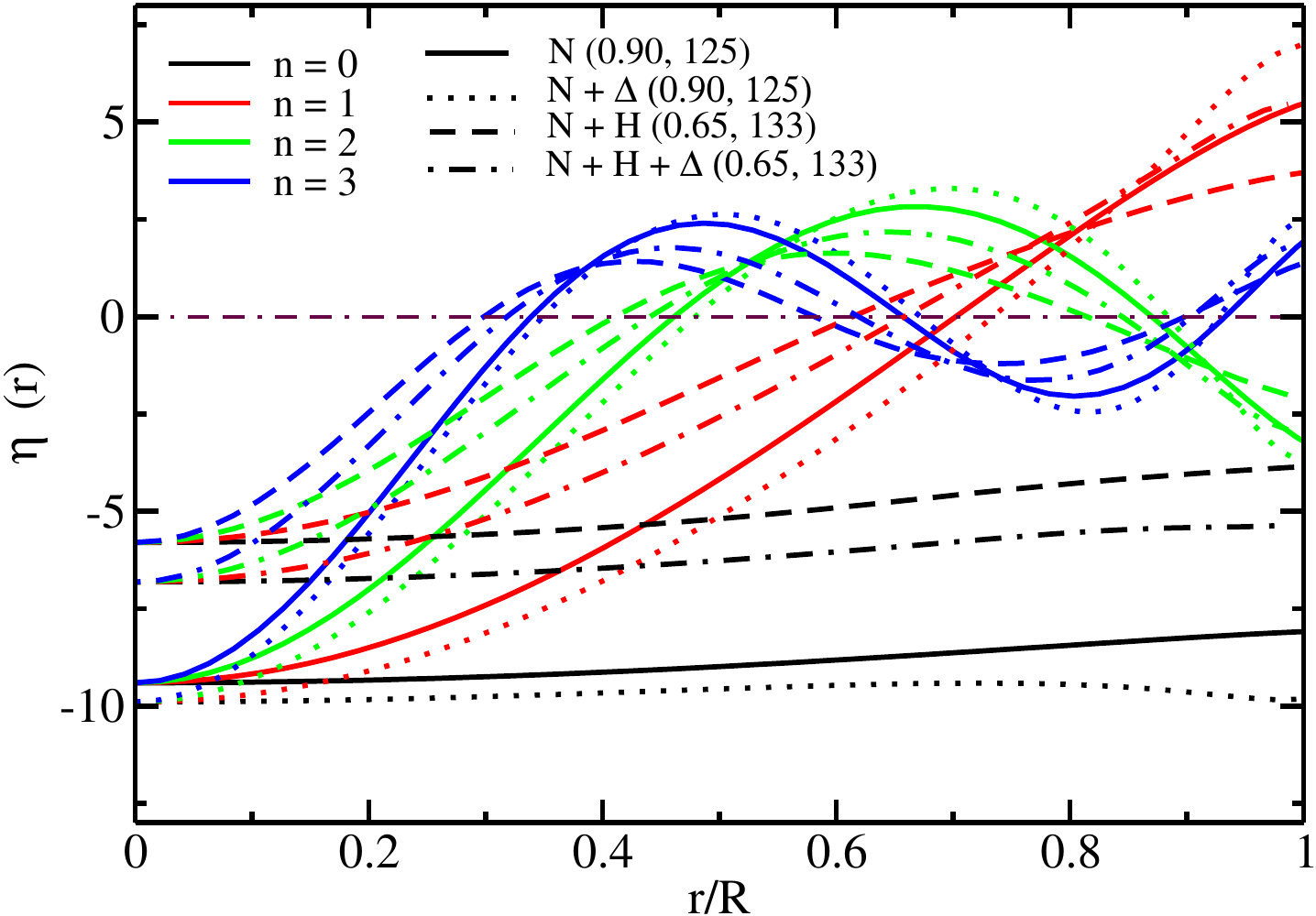}
\caption{The radial displacement perturbation ${\xi(r)}$ = ${\Delta r/r}$ (right panels) and the radial pressure perturbation ${\eta(r)}$ = ${\Delta r/r}$ (left panels) as a function of dimensionless radius distance ${r/R}$ for lower ${f}$-mode (n = 0), lower order ${p}$-modes (n = 1, 2, 3). The upper panels represent the result for N (solid lines), N+${\Delta}$ (dotted lines), N+H (dashed lines), and N+H+$\Delta$ (dot-dashed lines) hybrid EoS at 1.4 ${M_{\odot}}$ with quark parameter values (${C, D^{1/2}}$) = (0.90, 125 MeV) for N and N+${\Delta}$ EoS and  (${C, D^{1/2}}$) = (0.65, 133 MeV) for N+H and N+H+${\Delta}$ EoS. The lower panels represent the result for the same configuration of the EoSs at 1.8 ${M_{\odot}}$.}
\label{fig:radial}
\end{figure*}

 The radial displacement perturbation profile, ${\xi(r)}$, and pressure perturbation profile, ${\eta(r)}$, as a function of the dimensionless radius distance, ${r/R}$, are plotted in Figure \ref{fig:radial} at 1.4 ${M_{\odot}}$ (upper panels) and 1.8 ${M_{\odot}}$ (lower panels). Those profiles are plotted for hybrid EoS with four different particle compositions, N (solid lines), N+$\Delta$ (dotted lines), N+H (dashed lines), and N+H+$\Delta$ (dot-dashed lines). Only the ${f}$-mode (n = 0) and three excited modes (n = 1, 2, 3) are shown. 

Throughout the stars, in the range ${0 \leq r \leq R}$, exactly ${n}$ nodes are obtained for the ${n}$th mode, both for ${\xi}$ and ${\eta}$ profiles, as expected for any Sturm-Liouville system. This is clearly seen in the right panels for the $\eta$ radial profile, although in the $\xi$ case the figures become crowded due to displaying different models and modes. From Figure \ref{fig:radial} (left plots), one can see that the amplitude of ${\xi_n(r)}$ for each frequency mode ${\nu_n}$ is larger near the center and small at the surface. The lower modes show a smooth drop in their profiles while the higher modes depict small oscillations which would become large for higher modes. The fundamental profile, $\xi_0$, for N+${\Delta}$ and N+H+${\Delta}$ show an increase in the amplitude near the surface as compared to the N and N+H profiles. At 1.8 $M_{\odot}$, while the $\xi_0$ profile for N+${\Delta}$ shows an increase in the amplitude, the profile for N+H+${\Delta}$ decreases at first before increasing by a small amount near the surface. For the N+H EoS, the fundamental profile starts decreasing at around 0.6R. One can see a rapid sign change near the center of the star which along with the amplitude decreases as one moves toward the surface of the star. From Figure \ref{fig:radial} (right plots), we observe that the amplitude of ${\eta_n(r)}$  is larger near the center and also at the surface of the star. The ${\eta}$ profiles are more compact at 1.4 $M_{\odot}$ as compared to that for 1.8 $M_{\odot}$. Also, the fundamental radial pressure profile, ${\eta_0}$, at 1.4 $M_{\odot}$, tends towards a large negative amplitude near the surface for almost all hybrid EoSs. For the same profiles at  1.8 $M_{\odot}$, we can see that the amplitude starts decreasing. In our previous work without phase transition \cite{PhysRevD.107.123022}, we observed some different behavior for radial pressure and radial displacement profiles when the $\Delta$ baryons and hyperons appeared. With these exotic baryons and a slow phase transition to the quark matter, the interplay between these particles causes the fundamental profiles to behave quite differently at different star masses. Even though the ${\eta}$ oscillations are directly proportional to the Lagrangian pressure variation ${\Delta P}$, the amplitude of ${\eta_n(r)}$ for consecutive ${n}$ have large amplitudes near the surface, and thus the contribution from consecutive modes, ${\eta_{n+1}}$ - ${\eta_n}$ cancels out because of the opposite signs, thereby satisfying the boundary condition that ${P(r = R)}$ = 0. This implies that ${\eta_{n+1}}$ - ${\eta_n}$  and also ${\xi_{n+1}}$ - ${\xi_n}$ are more sensitive to the star's core. As a result, the measurement of ${\Delta \nu_n}$ = ${\nu_{n+1}}$ - ${\nu_n}$ is an observational imprint of this star's innermost layers.

In our investigation of radial profiles for SSHS, we found that the profiles closely resemble those observed in traditionally stable NSs. Due to the appearance of SSHS at the higher mass configurations,  we observed a slight reduction in both the amplitude and frequency of the radial modes for SSHS. This reduction is consistent with the expectations due to the higher mass and unique composition of SSHS, likely influenced by the presence of strange quark matter. However, qualitatively, the similarity in the radial profiles between SSHS and NSs suggests underlying structural similarities, albeit with nuanced differences reflecting the distinct nature of SSHS configurations.

\begin{table}[ht]
\centering
		\caption{10 lowest order radial oscillation frequencies, ${\nu}$ in (kHz) for N and N+${\Delta}$ EoS at quark parameter values (${C, D^{1/2}}$) = (0.90, 125 MeV), and N+H and N+H+${\Delta}$ EoS at quark parameter values (${C, D^{1/2}}$) = (0.65, 133 MeV). For each EoS, the frequencies are calculated at 1.4 ${M_{\odot}}$ and 1.8 ${M_{\odot}}$ of the corresponding star.\label{table1} }
\begin{tabular}{ p{1.5cm}p{1.5cm}p{1.5cm}p{1.5cm}p{1.5cm} }
 \hline
\multirow{2}{*}{Nodes} &\multicolumn{4}{c}{EoS} \\
 \cline{2-5}
  & N&N+$\Delta$ & N+H & N+H+$\Delta$\\
 \hline
 & &  1.4 $M_{\odot}$ & & \\
\hline 
0   &3.109  &2.591  &3.109	&2.504\\
1	&9.262	&8.463	&9.244	&8.030\\
2	&15.527	&14.766	&15.470	&14.095\\
3	&21.786	&20.927	&21.697	&20.080\\
4	&28.038	&26.994	&27.926	&25.814\\
5	&34.287	&32.993	&34.117	&31.719\\
6	&40.532	&38.934	&40.320	&37.379\\
7	&46.776	&44.824	&46.567	&43.141\\
8	&53.021	&50.676	&52.787	&48.836\\
9	&59.263	&56.507	&58.982	&54.453\\
\hline 
& &  1.8 $M_{\odot}$ & & \\
\hline 
0   &2.858  &2.465  &2.042	&2.097\\
1	&7.926	&7.404	&6.736	&6.557\\
2	&13.023	&12.604	&10.872	&10.827\\
3	&18.170	&17.877	&15.422	&15.551\\
4	&23.332	&23.117	&19.695	&20.168\\
5	&28.500	&28.317	&23.984	&24.612\\
6	&33.671	&33.483	&28.531	&29.180\\
7	&38.843	&38.618	&32.782	&33.789\\
8	&44.016	&43.724	&37.132	&38.231\\
9	&49.188	&48.803	&41.616	&42.675\\
 \hline
\end{tabular}
\end{table}

\begin{figure*}[h]
\centering
    \includegraphics[width=0.47\linewidth]{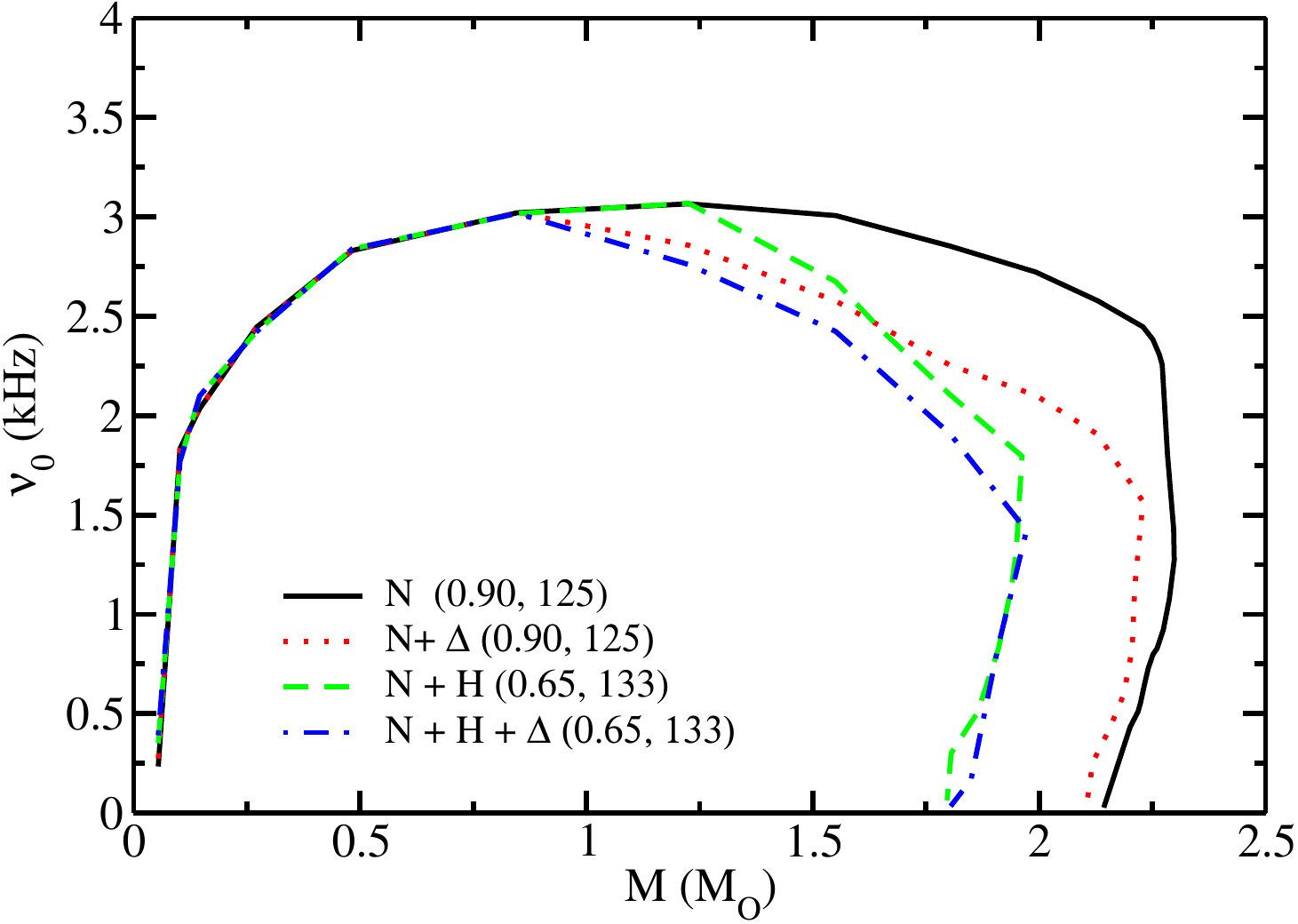}\hfil
.    \includegraphics[width=0.47\linewidth]{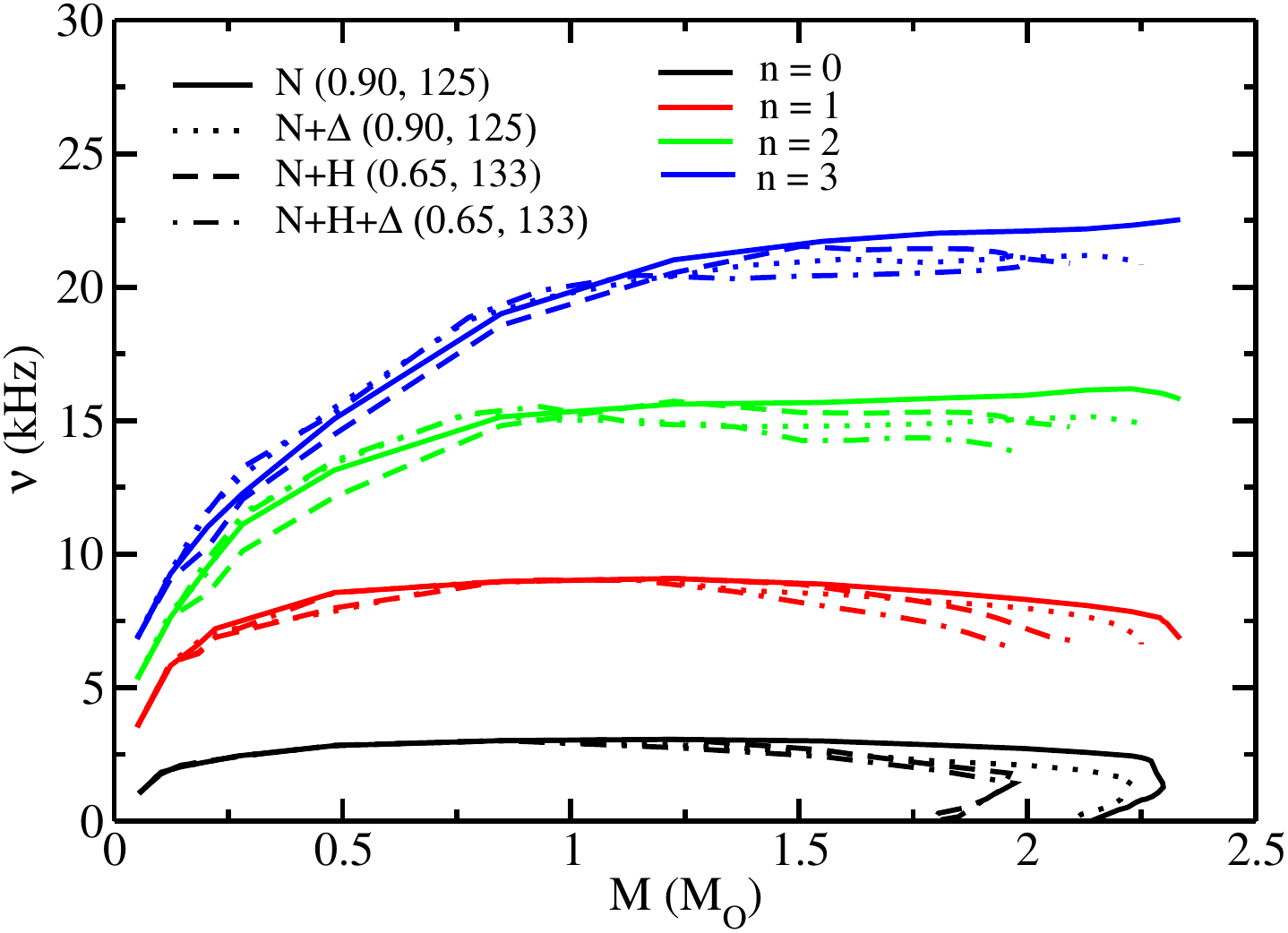}
	\caption{Fundamental frequencies (left panel) and fundamental frequency along with higher order frequencies of radially oscillating hybrid stars as a function of the mass sequence for N (solid lines), N+${\Delta}$ (dotted lines), N+H (dashed lines), and N+H+${\Delta}$ (dot-dashed lines) EoS at parameter values ($C, D^{1/2}$) = (0.90, 125 MeV) for N and N+${\Delta}$ EoS and  (${C, D^{1/2}}$) = (0.65, 133 MeV) for N+H and N+H+$\Delta$ EoS. The results are shown for the configurations at 1.4 ${M_{\odot}}$.}
	\label{fig7} 
\end{figure*}

Figure \ref{fig7} shows the frequencies of radially oscillating NSs with four different matter compositions, as a function of the stellar mass, for the four lowest radial modes, $n=0,1,2,3$ (right panel) and for the fundamental mode only (left panel). In the left panel of Figure \ref{fig7}, we see that the fundamental frequency initially increases with the mass and then starts decreasing at around 1 $M_{\odot}$. For the hybrid EoS with N+$\Delta$, the early appearance of $\Delta^-$ baryon shifts the frequency to lower values. For the hybrid EoS with N+H and N+H+$\Delta$, the frequency shift to lower values takes place at around 1.2 and 0.9 ${M_{\odot}}$, respectively. In general case, as the center density inside the star increases, the $f$-mode frequency starts to shift toward zero at the same time, the star is approaching its stability limit. The stability limit itself exhibits an eigenmode with zero frequency. On the contrary, maximum mass is reached for slow phase transitions before the fundamental mode's frequency vanishes indicating that some stellar configurations with higher central densities than maximum mass remain stable even with small radial perturbations. The right panel of Figure \ref{fig7} shows the fundamental as well as some higher modes as a function of mass for the hybrid star sequence for the same stellar configurations. Here, we observe the emergence of different exotic baryons as well as a phase transition to the deconfined quark matter. The increase in kink count in higher-order modes causes the higher-order frequencies to behave differently from fundamental frequencies. This represents an essential finding that results in a sequence of "avoided crossings" between the different modes: as two subsequent modes from different families approach one another, their frequencies reject each other \cite{1997A&A...325..217G,kokkostas}. This "avoided crossing" is evident at lower densities in all four cases and is a feature of a realistic EoS \cite{kokkostas}. Because there are a lot of exotic baryons as well as quarks in the system for higher-mass stars in our case, the kinks are more noticeable. Comparing these results with our earlier work \cite{PhysRevD.107.123022}, where we studied the radial oscillations of different baryonic compositions without a phase transition, we observed a different behavior of fundamental frequency. The appearance of exotic baryons at higher densities was observed clearly with the shift in the $f$-mode frequency. Although the frequencies of subsequent modes from different families rejected each other, the kinks present were small as compared to this study. Thus, the appearance of different phases in the high-density nuclear matter completely alters the fundamental and overtone frequencies of the NS sequence. As a result, measurements of stars' radial oscillations can reveal more about the microphysics of the NS interior.

Table \ref{table1} shows the frequencies ${\nu}$ of the first 10 nodes in kHz for hybrid EoSs with pure nucleonic matter N, hypersonic matter N+H, and ${\Delta}$-included nucleonic and hyperonic matter N+$\Delta$ and N+H+$\Delta$, respectively. The frequencies are obtained at 1.4 $M_{\odot}$ and 1.8 $M_{\odot}$ to observe the difference between the frequencies at two different mass values. The node ${n}$ = 0 corresponds to the ${f}$-mode frequency, while the others correspond to the excited ${p}$-modes. The frequency for the ${f}$-mode and other higher modes are small at 1.8 $M_{\odot}$ when compared to the values at 1.4 $M_{\odot}$. The hybrid EoS with nucleons only predicts higher values of the frequency for all modes compared to other hybrid EoS with $\Delta$s and hyperons.

\begin{figure}[ht]
\centering
	\includegraphics[scale=0.35]{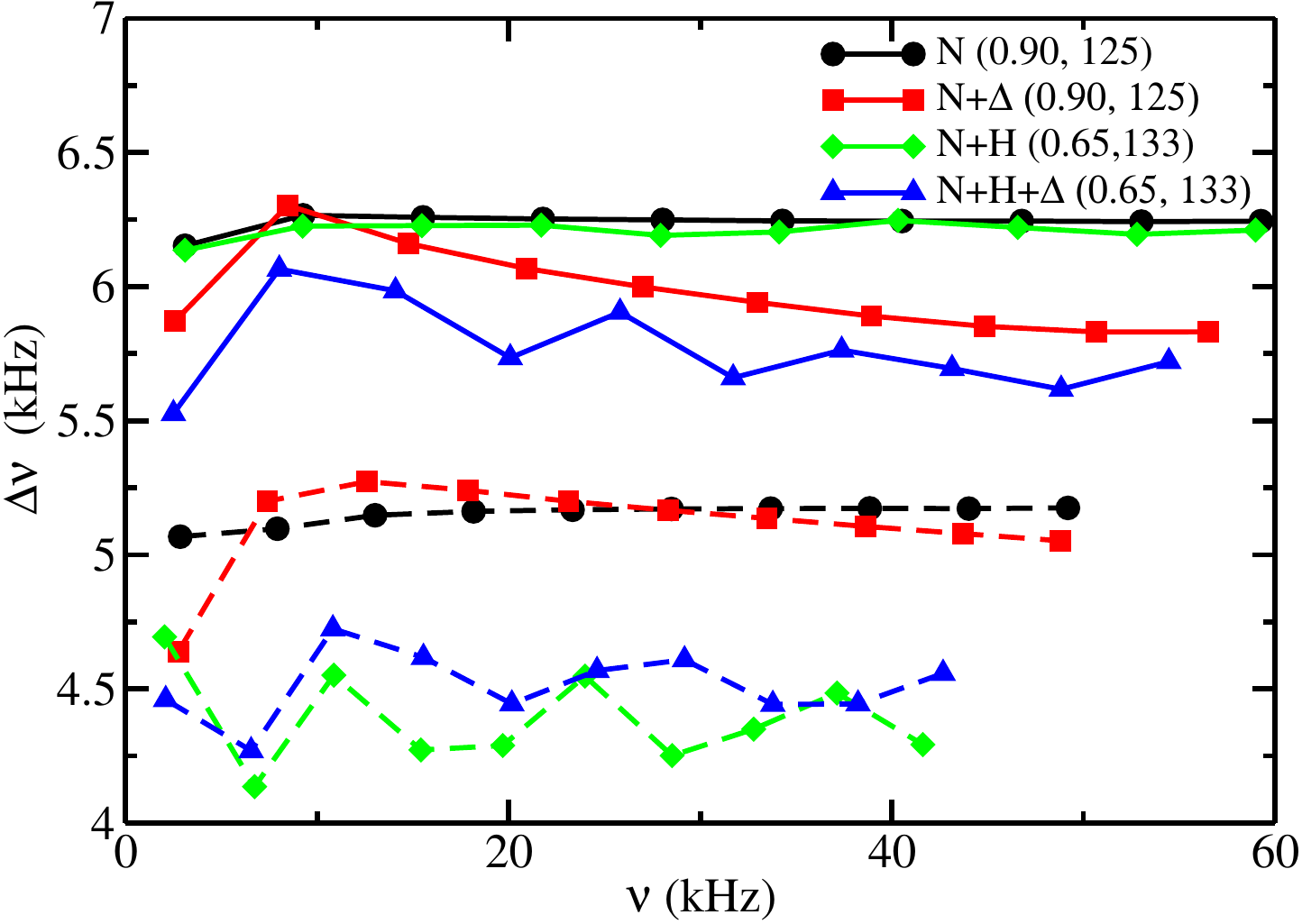}
	\caption{Frequency difference ${\Delta \nu_n}$ = ${\nu_{n+1}}$ - ${\nu_n}$ versus ${\nu_n}$ in kHz for N and N+${\Delta}$ hybrid EoS at parameter values (${C, D^{1/2}}$) = (0.90, 125 MeV), and for N+H and N+H+${\Delta}$ hybrid EoS at (${C, D^{1/2}}$) = (0.65, 133 MeV). Solid lines represent the frequency difference at 1.4 ${M_{\odot}}$ and dashed lines correspond to values at 1.8 ${M_{\odot}}$.}
	\label{fig8} 
\end{figure}

A difference between consecutive modes in asteroseismology, referred to as the large separation, or ${\Delta \nu_n}$ = ${\nu_{n+1}}$ - ${\nu_n}$, is frequently used to interpret the properties of stars \cite{PhysRevD.101.103003, PhysRevD.101.063025, PhysRevD.107.123022}.
Figure~\ref{fig8} displays the frequency difference ${\Delta \nu_n}$ = ${\nu_{n+1}}$ - ${\nu_n}$ vs ${\nu_n}$ in kHz for hybrid EoSs with N, N+${\Delta}$, N+H, and N+H+${\Delta}$ matter. The large separation for each Eos is calculated at 1.4 (solid lines) and 1.8 ${M_{\odot}}$ (dashed lines). For the hybrid EoS with N and N+H at 1.4 ${M_{\odot}}$, the separation between the modes is almost the same and there are no fluctuations at lower modes (${n}$ = 0).  For the hybrid EoS with N+$\Delta$, we see that the separation between the modes decreases while for N+H+$\Delta$, we observe some.  Similarly at 1.8 ${M_{\odot}}$, the hybrid EoS with N (N+$\Delta$) shows almost no (small) change in the frequency difference between consecutive modes.  For N+H and N+H+$\Delta$ hybrid EoS, we observe the signature fluctuations present in ${\Delta \nu_n}$. These fluctuations arise from the significant variation of the speed of sound squared ${c_s^{2}}$ or the relativistic adiabatic index ${\gamma}$ on the transition layer separating the inner and outer core of the NS, which has an amplitude proportionate to the magnitude of the discontinuity. The reason for not observing any fluctuations in some hybrid EoS could be due to the fact that we have used a non-unified EoS in the present study, which is different in comparison to our previous study without a phase transition \cite{PhysRevD.107.123022}. Another reason could be the interplay between different exotic baryons and a phase transition to the deconfined quark matter.
However, the radial oscillation for the lowest order mode (${n}$ = 0) is not greatly impacted by the crust because it typically accounts for less than 10\% of the stellar radius and the oscillation nodes are situated far inside the NS core. But other high oscillation modes are present in the crust of the star and hence the eigenfrequencies are modified (characterized by the peaks in the ${\Delta \nu_n}$) \cite{Sen:2022kva, PhysRevD.107.123022}.

\section{Summary and Conclusion}
\label{summary}

To summarize our work, we have extended our previous study on the radial oscillations of neutron and hyperon stars with exotic ${\Delta}$ baryons by allowing a phase transition to the deconfined quark matter, employing the density-dependent quark mass (DDQM) model. For the hadronic matter, we use the previously employed DDME2 parameter set from the DDRMF model and describe the spin-${3/2}$ ${\Delta}$ baryons using the Rarita-Schwinger Lagrangian density. We have used the Maxwell construction to produce hybrid stars, and the hadron-quark phase transition is assumed to be first-order. It is necessary to keep in mind that phase conversions close to the interface can result from small radial perturbations in hybrid stars with sharp discontinuities. The speed at which the hadron-quark conversion reactions occur determines the dynamic stability of hybrid stars. There are two limiting cases: slow and rapid conversions, depending upon the conversion timescales. Necessary junction conditions are required and used for the numerical integration of the oscillation equations in hybrid stars. We studied the hadron-quark conversion with a slow phase transition, as this allows us to study the possibility of a new class of stable hybrid stars known as slow stable hybrid stars (SSHS).

After determining the coexistence point, we observe that the EoSs reach a plateau at the coexistence pressure in intermediate energy densities and then proceed to rise as the deconfined EoS following the phase transition. Throughout the whole energy density range, the pure nucleonic EoS is no longer the stiffest. Instead, because the coexistence point is under more pressure at intermediate energy density values, the N+$\Delta$ EoS becomes the stiffest. Also, the presence of $\Delta$s allows for a delayed phase transition, moving the coexistence point to higher densities as can be seen in the population plots. The variation of the adiabatic index, $\gamma$, with respect to the energy density is studied for the different hybrid EoSs constructed. For hybrid EoSs with N and N+H, $\gamma$ increases to a peak value and proceeds to drop smoothly for the N EoS and sharply for the N+H EoS before the phase transition to quark matter causes the adiabatic index to drop to zero. For EoSs with N+$\Delta$ and N+H+$\Delta$, the adiabatic index attains higher values for most of the energy density range because of the delayed phase transition to the deconfined quark matter. The kinks represent the appearance of new particles. After the phase transition, $\gamma$ presents an almost constant value for pure quark matter.

$c_s^2$ drops to zero at the coexistence point due to the phase transition to quark matter, increasing again for pure quark matter. The kinks present correspond to the presence of different particles before the phase transition and are visible for hybrid EoS with different particle compositions. The conformal limit is breached at lower densities by the pure baryonic part of $c_s^2$ for hybrid N, N+$\Delta$, and N+H+$\Delta$ EoS. We observe a high $c_s^2$ at intermediate densities for the hybrid N+H+$\Delta$ composition curve, due to the early appearance of $\Delta^-$ and the delayed phase transition to quark matter. Ultimately, all four speeds of sound approach the conformal limit from below at high energy densities.

Regarding hydrostatic equilibrium, we integrated the structure equations numerically to obtain the mass-radius relation for hybrid stars with different hadronic compositions. Although the maximum mass decreases from 2.30 $M_{\odot}$ for hybrid N EoS to 2.25 $M_{\odot}$ for hybrid N+$\Delta$ EoS, the change in radius at both the maximum and canonical masses is large, at 1.2 and 0.5 km respectively. For the hybrid EoS with hyperons and $\Delta$s, N+H and N+H+$\Delta$, the maximum mass increases from 1.97 to 1.98 $M_{\odot}$, while the radius at maximum and canonical mass decrease to 11.57 and 12.97 km, respectively. The appearance of hyperons along with deltas and the subsequent phase transition to quark matter soften the EoS, such that we obtain a star with a maximum mass of less than 2 $M_{\odot}$ despite our choice of a stiff quark EoS.
All hybrid models foresee a high transition density, especially for EoSs with $\Delta$s. The configurations to the left of the maximum mass star are also stable, called SSHS, since we applied the necessary junction conditions at the phase transition and utilized a slow phase conversion. In other words, the central densities of these stars exceed the central density of the star with $M_{max}$. The length of these SSHS branches is determined by the density jump between two phases and the stiffness of the quark EoS. For N+H hybrid EoS, the density jump $\Delta \mathcal{E}$ between the hadronic and quark phase is small among all hybrid EoSs, and hence the length of SSHS for this hybrid EoS is large.
 
To examine the radial oscillations of these pulsating stars, the equations for the perturbations are also solved. This allowed us to calculate the nine excited ${p}$-modes and the fundamental ${f}$-mode frequencies of the modes, as well as their corresponding eigenfunctions at 1.4 and 1.8 $M_{\odot}$, by solving the Sturm-Liouville boundary problem and imposing appropriate conditions at both the center and the surface of the stars. Our numerical results show that, in contrast to the N and N+H profiles, the fundamental profiles, $\xi_0$, for N+$\Delta$ and N+H+$\Delta$ at 1.4 $M_{\odot}$ exhibit an increase in amplitude near the surface. At 1.8 $M_{\odot}$, the $\xi_0$ profile for N+$\Delta$ exhibits an increase in amplitude, whereas the profile for N+H+$\Delta$ initially decreases and then slightly increases near the surface. At 1.4 $M_{\odot}$, the $\eta$ profiles are more compact than those at 1.8 $M_{\odot}$. Furthermore, at 1.4 $M_{\odot}$ the fundamental radial pressure profile, $\eta_0$, tends towards large negative amplitudes near the surface. The amplitude begins to decrease for the same profiles at 1.8 $M_{\odot}$. The fundamental profiles exhibit distinct behaviors at different star masses due to the interplay between exotic baryons and a slow phase transition to quark matter. Additionally, we observed that, for slow phase transitions, the maximum mass is reached before the frequency of the fundamental mode vanishes, suggesting that some stellar configurations with higher central densities than the maximum mass are stable even in the presence of small radial perturbations.
We have also studied the radial profiles for the SSHS, and we have found that the SSHS eigenmodes do not show (qualitative) differences in comparison to the traditionally stable neutron stars, as the profiles closely resemble each other. Due to the appearance of SSHS at the higher mass configurations, we observed low values of both the amplitude and the frequency of the radial modes for SSHS. As future work, it would be interesting to compute the radial profiles for the whole M-R curve, to see whether or not we can identify features that may differentiate between the SSHS and traditionally stable NSs.

We also studied the frequency difference between consecutive modes and observed that, for the hybrid EoS with N and N+H at 1.4 $M_{\odot}$, the separation between the modes remains almost unchanged and there are no fluctuations. With $\Delta$ baryons, the separation between modes decreases. For the same calculation at 1.8 $M_{\odot}$, we observe erratic fluctuations for hybrid EoS with hyperons and $\Delta$s.

In conclusion, radial oscillations of hybrid stars under a slow phase conversion present a fascinating and intricate exploration into the fundamental aspects of nuclear matter. The study delves into the dynamic processes occurring during the transition between hadronic and quark phases, shedding light on the behavior of hyperons and delta baryons within this complex framework. Further work on radial oscillations in newborn NSs after supernova explosions or NS mergers should also consider more realistic characteristics of these environments, such as temperature, rotation, and magnetic field, for a more detailed exploration of the intricate dynamics involved in slow phase conversions.

Advances in multi-messenger astronomy and third-generation ground-based gravitational wave detectors, such as Einstein Telescope and Cosmic Explorer, complemented by robust microscopic models enabling exploration of various neutron star compositions, including hyperon content, are anticipated to impose precise limitations on the equation of state of baryonic matter under high-density conditions. \cite{Providencia:2023rxc,Tsang:2023vhh}.

\acknowledgments

We would like to extend our sincere gratitude to the anonymous reviewer for invaluable contribution to the enhancement of our manuscript which has elevated the quality of our research.
I. A. R. acknowledges support from the Alexander von Humboldt Foundation. K. D. M. was supported by the Conselho Nacional de Desenvolvimento Científico e Tecnológico (CNPq/Brazil) under grant 150751/2022-2. B. C. B. was supported by STFC through her PhD studentship under grant ST/W50791X/1. I. L. acknowledges the Funda\c c\~ao para a Ci\^encia e Tecnologia (FCT), Portugal, for the financial support to the Center for Astrophysics and Gravitation (CENTRA/IST/ULisboa) through the grant Project~No.~UIDB/00099/2020 and grant No. PTDC/FIS-AST/28920/2017.



\providecommand{\href}[2]{#2}\begingroup\raggedright\endgroup


\end{document}